\documentclass[aps,pra,10pt,twocolumn,superscriptaddress]{revtex4}
\usepackage{graphicx}
\usepackage{amssymb, amsmath}
\usepackage[usenames, dvipsnames]{color}
\usepackage{ulem}
\usepackage{graphicx}
\usepackage{subfig}
\usepackage{dcolumn}
\usepackage{latexsym}
\usepackage{amsbsy}
\usepackage{bbold}
\usepackage{mathtools}
\usepackage{enumitem}
\usepackage[version=3]{mhchem}

\usepackage{ulem}

\def\I{\uppercase\expandafter{\romannumeral 1}}
\def\II{\uppercase\expandafter{\romannumeral 2}}
\def\III{{\uppercase\expandafter{\romannumeral 3}}}
\def\IV{{\uppercase\expandafter{\romannumeral 4}}}
\def\V{{\uppercase\expandafter{\romannumeral 5}}}
\def\VI{{\uppercase\expandafter{\romannumeral 6}}}
\def\VII{{\uppercase\expandafter{\romannumeral 7}}}
\DeclareMathAlphabet{\mathsfsl}{OT1}{cmss}{m}{sl}

\DeclareMathOperator{\Sgn}{Sgn}
\DeclareMathOperator{\erfc}{erfc}

\newcommand\inv[1]{#1\raisebox{0.3em}{\tiny-\!1}}
\newcommand\invbig[1]{#1\raisebox{0.6em}{$\scriptscriptstyle-\!1$}}

\numberwithin{equation}{section}

\begin{document}
\title{ Quench Dynamics of the Gaudin-Yang Model}

\author{Huijie Guan and
   Natan Andrei\\ Department  of  Physics,  Rutgers  University,
Piscataway,  New  Jersey  08854 }

\begin{abstract}
We study the quench dynamics of one dimensional  bosons or fermion quantum gases with either attractive or repulsive contact interactions. Such systems are well described by the Gaudin-Yang model which turns out to be quantum integrable. We use a contour integral approach, the Yudson approach, to expand initial states in terms of  Bethe Ansatz eigenstates of the  Hamiltonian. Making use of the contour, we obtain a complete set of eigenstates, including both free states and bound states. These states constitute a larger Hilbert space than described by the standard String hypothesis.  We calculate the density and noise correlations of several quenched systems such as a static or kinetic impurity evolving in an array of particles. 
\end{abstract}
\maketitle

\section{Introduction}
The study of nonequilibrium dynamics has been stimulated recently by the remarkable progress in the ultracold atom systems and other systems that allow fine control of its parameters and isolation from the environment. These systems provide us with finely engineered model Hamiltonians, arbitrarily designed initial state and single-site and single-atom resolution. This developments spurred many questions, such as whether steady states emerge, how do observables equilibrate or validity of thermodynamic ensembles in describing the equilibrium states in large isolated systems.

We shall study  two-component  atomic gases  in one dimension interacting via short range potentials, the atoms being either of fermions or of bosons. The system may be realized experimentally using $\ce{^{6}Li}$~\cite{liao2010spin} or $\ce{^{40}K}$~\cite{moritz2005confinement} atoms loaded in a one dimensional optical lattice. Millions of spin-polarized fermion atoms can be cooled well below the Fermi temperature. Then a desired spin population imbalance can be prepared by Laudau-Zener radio frequency sweep. Moreover, the kinetic part of the Hamiltonian may be controlled by the optical lattice while the Feshbach resonance may tune the interaction strength. Lastly, the density profile for each spin state can be measured by state-selective imaging procedure with down to single site and single atom resolution.

 These systems are described by the Gaudin-Yang model which is exactly solvable by the Bethe Ansatz.  This allows a complete determination of its thermodynamic properties \cite{guan2013fermi} using the thermodynamic Bethe Ansatz (TBA) approach \cite{takahashi1994one}. It also allows an exact study of many of its non equilibrium properties, some of which will be studied in this paper.  The study is carried out via the Yudson approach based on contour integral representation that solves time-evolution of one dimensional integrable systems analytically.   So far it has been used to solve the quench dynamics in the Lieb-Liniger gas~\cite{deepak1, deepak2} and the XXZ model~\cite{liuws}, beyond the original work by Yudson.  In this paper, the  approach will be  generalized for the Gaudin-Yang model with 
nested Bethe Ansazt. It includes two sets of integral contour, one for quasi-momenta and one for spin rapidities.

  The  Yudson approach  proceeds by expanding the  initial state  in energy eigenstates  with their overlaps with the initial state  being calculated 
  with respect to   ordered Yudson states and integrated 
  over appropriately chosen contours that capture the poles of the S-matrices  and render the representation exact.  It  completeness is formally proven and does not rely on  the String hypothesis which states that the parameters characterizing the state, real or imaginary, follow some well defined  patterns - "strings", see below. Rather,  the Yudson approach allows to deduce the hypothesis  directly in the infinite volume limit. The string hypothesis has been widely used and  leads in general to excellent results, but has not been proven and in some instances can be shown to be incomplete \cite{destri1982analysis}.  In this work, we found string structures that are beyond the hypothesis and some of the strings predicted in the conjecture need to be modified.

Other approaches to  quench dynamics of integrable models include  the Quench Action method~\cite{caux2016quench} and ABACUS  (\underline{A}lgebraic \underline{B}ethe \underline{A}nsatz-based \underline{C}omputation of \underline{U}niversal \underline{S}tructure factors)~\cite{Abacus1}. The question of quench dynamics has been widely studied   in many other context and other methods while  including:  t-DMRG~\cite{feiguin2005finite, essler2014quench,schwarz2017nonequilibrium}, exact diagonalization~\cite{zangara2013time,zhang2011quantum}, 
t-RG \cite{kashuba2013quench}, Flow equations \cite{heyl2010interaction}, TEDB \cite{Mines}

We calculate the density and noise correlation from several initial states such as a static or kinetic impurity evolving in an array of particles. Among our observations: starting from a  Mott state with small overlap among the particles the system retains this feature after the quench. This is reflected in the density evolution which displays similar behavior for systems with different signs of interaction, indicating the suppression of bound states. At the same, the study of normalized noise function $c(z,-z)$ at the origin shows different stages. Shortly after the quench, the sign of the correlation function $c(0,0)$ depends on whether the interaction is attractive or repulsive interaction. subsequently  $c(0,0)$ quickly approaches the value where the possibility to find both particles is small, later as time goes on, the value of the correlation function at origin increase gradually for attractive models while remaining small for repulsive systems. Other properties that appear relate to the FFLO type states that appear when there is an unequal number of the two components.  Such states involve singlet pairs with nonzero center of mass momentum that equals the mismatch of the two Fermi sea. The exotic phase manifests itself as oscillatory order parameter as compare to the homogeneous BCS phase~\cite{feiguin2012bcs}. Unlike in the low energy scenario where the FFLO phase occupies a lot of the parameter space in 1D, the FFLO state is absent in the quench dynamics in our study, which involves numerous amount of excitations. Even though we observed dynamical emergence of singlet bound states, they are suppressed due to minimum overlap with the initial state.

The structure of the paper is as follows. In section 2, we review existing Bethe Ansatz results for the Gaudin-Yang model, which are the building blocks for the following sections. Section 3 is denoted to the introduction of the Yudson approach where advantages and disadvantages are discussed. Explicit form of the Yudson representation for the Gaudin-Yang model is given in section 4. And integration contours are specified for both attractive and repulsive cases. In section 5, we separate out bound states from the contour and compare the string solutions with the String hypothesis.
In section 6, we calculate the time evolution of a single impurity problem. We first obtain the exact wavefunction for two particle cases and derived the multi-particle wavefunction in the asymptotic limit. In section 7, we repeat the calculation for bosonic Gaudin-Yang model and make comparisons between the two systems. In the last section, we summarize our work and discuss promising directions with the approach.

%attract a lot of theoretical interests. One fascinating property of these systems is existence of exotic pairing state, which is called the Fulde-Ferrell-Larkin-Ovchinnikov (FFLO) state. This bound state consists of opposite spins with nonzero center of mass, which presents itself as oscillatory order parameter~\cite{guan2013fermi}. 

\section{The Fermionic Gaudin-Yang Model}\label{sec:GYM}
The  model describes two-component Fermi gases with contact interaction confined in a single dimension. The Hamiltonian is defined as
\begin{align}\label{HGY}
&H_{GY}=\notag\\
&\sum_{\sigma=\uparrow,\downarrow}\int_x \partial_x\Psi^\dagger_\sigma(x)\partial_x\Psi_\sigma(x)
+c\int_x\Psi^\dagger_{\uparrow}(x)\Psi^\dagger_{\downarrow}(x)\Psi_{\downarrow}(x)\Psi_{\uparrow}(x)
\end{align}
$\Psi^\dagger_\sigma(x)$ ($\Psi_\sigma(x)$) is the creation (annihilation) operator of a spin $\sigma$ fermion at location $x$, $c$ characterizes the interaction, which can be attractive($c<0$) or repulsive($c>0$). The model is integrable and can be solved by the nested Bethe Ansatz. The solution for arbitrary spin population imbalance was first obtained by Yang~\cite{PhysRevLett.19.1312} and Gaudin~\cite{gaudin1967systeme}. The eigenstates are characterized by quasimomenta $k$'s and spin rapidities $\mu$'s. For a system with $N-M$ majority fermions of one spin and $M$ fermions of the opposite spin, they take the form
\begin{align}\label{Est}
|\mu,k\rangle=&\sum_{P\in S_N,R\in S_M}\int_x\sum_{\alpha} (-1)^P e^{i(Pk)_{i}x_i}\prod_{i<j}S(\mu_i-\mu_j)\notag\\
&\prod_{i=1}^M
I(\mu_i,Pk,\alpha_{\inv{R}i})\theta(\alpha)\theta(x)|x,\alpha\rangle
\end{align} 
with
\begin{align}\label{S_mu}
S(\mu_i-\mu_j)=\frac{\mu_i-\mu_j+ic\Sgn(\alpha_{
\inv{R}i}-\alpha_{\inv{R}j})}{\mu_i-\mu_j-ic}
\end{align}
\begin{align}\begin{split}
I(\mu,k,\alpha)=\frac{ic}{\mu-k_{\alpha}+ic/2}\prod_{n<\alpha}\frac{\mu-k_n-ic/2}{\mu-k_n+ic/2}
\end{split}\end{align}
%\begin{align}\begin{split}
%|x,\alpha\rangle=\prod_{i=1}^M
%\sigma_{\alpha_i}^- \prod_{j=1}^N\Psi_{\uparrow}^\dagger(x_i)|0\rangle
%\end{split}\end{align}
$P$, $R$ are permutation operators and $k_{Pi}=(\invbig{P}k)_i$. $x$'s are the locations of the fermions, $\alpha$'s are the labels of down spins along the chain. That is to say $|x,\alpha\rangle=\prod_{i=1}^N \Psi^\dagger_{a_i}(x_i)|0\rangle$ and $a_i=\downarrow$ if $i\in \{\alpha\}$, $a_i=\uparrow$ if $i\not\in\{\alpha\}$. $\theta(\alpha)$ is a shorthand notation, defined as $\theta(\alpha_1<\ldots<\alpha_M)$ and similarly, $\theta(x)=\theta(x_1<\ldots<x_N)$. The state satisfies the equation
\begin{align}
H|\mu,k\rangle=\sum_{i=1}^N k_i^2|\mu,k\rangle
\end{align}

The Bethe Ansatz equations are obtained by placing the systems under the periodic boundary condition. This leads to
\begin{align}\label{bethe1}
e^{ik_nL}=\prod_{j=1}^M\frac{k_n-\mu_j+ic/2}{k_n-\mu_j-ic/2}
\end{align}
\begin{align}\label{bethe2}
\prod_{j\neq i}\frac{\mu_i-\mu_j+ic}{\mu_i-\mu_j-ic}=\prod_{n}\frac{\mu_i-k_n+ic/2}{\mu_i-k_n-ic/2}
\end{align}

The solutions to these equations follow some pattern, as first proposed by Takahashi as three conjectures in~\cite{takahashi1994one}:
\begin{enumerate}
  \item Complex $\mu$ always form strings. For a $\mu$-string with length $n$, the $\mu$'s take the value $\mu_i=\mu+i(n+1-2j)c/2$ for $j=1,\ldots, n$.
  \item For $c>0$, momenta $k$'s are real.
  \item For $c<0$, complex $k$ forms complex conjugate pairs with another $k$ with a $\mu$ being its real part.
\end{enumerate}
These conjectures form the String hypothesis. They were  applied to other systems with internal degree of freedom, like the 1D Hubbard model~\cite{takahashi1972one}, the sine-Gordon model~\cite{fowler1981quantum} and the Kondo problem~\cite{andrei1983solution,tsvelick1983exact}. Although the hypothesis has not been proven explicitly, it is widely accepted and been used to obtain thermodynamic properties of these systems. However, as we will shown later in this work, we found that some type of strings predicted here does not show up, while other new type of strings do exist.  

Physically, these string solutions are related to bound states for attractive fermions, i.e. a $k$-pair describes a bound state between two fermions and a $\mu$-string depicts a bound state between down spins. It is easy to see that when $k=k_1+ic/2=$ $k_2-ic/2$, the wavefunction contains a factor \[e^{ik(y_{P1}+y_{P2})+c/2(y_{P1}-y_{P2})}\theta(y_{P1}-y_{P2})\] which decreases exponentially with the separation between $y_{P1}$ and $y_{P2}$. For states with $\mu=\mu_1+ic/2=$ $\mu_2-ic/2$, the wavefunction is proportional to \[\prod_{\alpha_{\inv{R}2}<m<\alpha_{\inv{R}1}}\frac{\mu-k_{\inv{P}m}}{\mu-k_{\inv{P}m}+ic}\prod_{n<\alpha_{\inv{R}2}}\frac{\mu-k_n-ic}{\mu-k_n+ic}\]
As $|\frac{\mu-k_{\inv{P}m}}{\mu-k_{\inv{P}m}+ic}|<1$ for real $k$ and $\mu$, the state diminishes with the distance between the two down spins.

One type of bound states of particular interest are the ones between a down-spin and an up-spin, i.e. between $k_{P\alpha_i}$ and $k_{Pj} (j\not\in \alpha)$. When the center of mass of the bound state is not zero, this is called a Fulde-Ferrell-Larkin-Ovchinnikov (FFLO) state which relates to the non-conventional superconducting state. Great efforts have been made to find the signature of such states in terms of local observables, both theoretically~\cite{lee2011asymptotic,hu2007phase} and experimentally~\cite{liao2010spin}. As shown in~\cite{lee2011asymptotic}, pair correlation and spin correlation function oscillate with wavefunction related to the spin population imbalance. The pair correlation in the momentum space shows peaks at the value of the mismatch between two fermi seas. In this paper, we will discuss how these bound state emerges dynamically as seen from the wavefunction.

\section{Yudson Approach}\label{ch: advantage}
The approach was originally  proposed by V. Yudson to study the superradiance effect in an infinite system described by the Dicke model in 1988~\citep{yudson1,yudson2}. The formalism has not gain much attention until recent years when the interest in nonequilibrium process has received a boost from the field of ultracold atoms. So far, the approach has successfully solved the quench dynamics of the Lieb-Liniger model~\citep{deepak1,deepak2} and XXZ model~\citep{liuws} on an infinite line, the Lieb-Liniger model with strong interaction on a finite line with periodic boundary conditions~\citep{garryLL} as well as hard wall boundary conditions~\citep{garryLLhardwall} and the Dicke model~\citep{colinDicke}. In this paper, we will use it to study the time evolution of two component fermion gases following a global quench.

The core of the Yudson approach is a resolution of the identity which takes the form
\begin{align}
\mathbb{1}=\int_{k,\mu}|k,\mu\rangle(k, \mu|
\end{align}

Here $|k,\mu\rangle$ represents the Bethe Ansatz eigenstate [see Eq. (\ref{Est})]  and $|k,\mu)$,  the Yudson state, is given by,
\begin{align}
\begin{split}
|\mu,k)=\int_x \sum_\alpha e^{ik_ix_i}\prod_{i=1}^MI(\mu_i,k,\alpha_i)\theta(\alpha)\theta(x)
|x,\alpha\rangle.
\end{split}
\end{align}
Namely, it is a single term in $|k,\mu\rangle$ corresponding to the identity permutation. The full Bethe Ansatz eigenstate can be expressed in terms of  the Yudson states,
\begin{align}\begin{split}
|\mu,k\rangle=&\sum_{P,R}S_{k,\mu}(P,R)|R\mu,Pk)
\end{split}\end{align}
with
\begin{align}\begin{split}
S_{k,\mu}(P,R)=(-1)^P\prod_{\substack{i<j\\ \inv{R}i>\inv{R}j}}\frac{\mu_i-\mu_j+ic}{\mu_i-\mu_j-ic}.
\end{split}\end{align}
Using the relation among  the eigenstate,
\begin{align}\begin{split}
|\mu,k\rangle=S_{k,\mu}(P,R)|R\mu, P k\rangle
\end{split}\end{align}
it is easy to see that the Yudson representation is nothing but the eigenstate expansion of the unity operator
$\mathbb{1}=\sum_{\mu, k}|\mu,k\rangle (\mu,k|\theta(\mu)\theta(k)$ with summations replaced by integrals. However, this modification results in significant simplifications in the non-equilibrium calculations, as listed below.

\begin{enumerate}
\item
It implifies the calculation of the overlap with the initial state.
\item
Obviates the need to solve Bethe Equations if one works on the infinite line.

\item
Complex contour includes the contribution from both free states and bound states (i.e. string states).
\item
Infinite rapidity guarantees that the expansions scans the whole Hilbert space, not only the highest weight states.
\item
Complicated norm factors disappear.
\end{enumerate}
We now discuss these points in turn. Point 1  stems from the simple form of a Yudson state. Instead of being a summation over factorially many terms, in Yudson representation, the overlap is one simple term.  Point  2 is related to the fact that the system is infinite while the Bethe equations originate from the constraint of  periodic boundary condition. Usually, the system is placed on a circle so as to make the momentum discrete. This is helpful to label the states and to include a few low energy states. Such boundary condition becomes unnecessary for the dynamics as all states overlapping with the initial state should be included. 
Point 3 is the most prominent. For systems  with bound states, the Hilbert space is spanned by states with complex parameters. Even though the roots are not scattered on the complex plane randomly, the number of string patterns they fall into is still huge. In Yudson approach, this difficulty is circumvented by a proper choice of integration contour in the complex plane. Such a contour integration incorporates contributions from both free states and bound states. To separate them apart, one simply shift all contours to the real axis.  
Point  4 is an interesting one. As  every nested Bethe Ansatz state is a highest weight state with respect to the total spin. This means that it will be annihilated by spin raising operator $S^+$. Therefore, a complete basis includes not only Bethe Ansatz states $|\mu, k\rangle$, but also states with lower expectation value of $S^z$, i.e. $(S^-)^n|\mu,k\rangle$ with $n=1, \ldots, N-2M$ for system with $N$ particles and $M$ down spins. However,  as shown in~\cite{braak2001spectrum}, the spin lowering operator simply corresponds to the spin wave with $\mu\to\pm \infty$. Therefore, by integrating from $-\infty$ to $+\infty$ for each rapidity, we are assured to probe every eigenstates in the Hilbert space. 
Point 5 is also associated with the infinite size of the system. In finite volume, the norm of an eigenstate is always complicated that consists of determinant whose dimension equals the number of degree of freedom~\citep{hubbardbook,quantuminversebook}. For Yudson representation, as long as the states are dimensionless, the norm is always a constant that does not depend on the system parameters. We will see this in the next section while proving the central theorem.

Note, the Yudson Approach does not depend on the eigenstates being complete and orthogonal, nor is it based on the String hypothesis. Instead, the proof of the Yusdon representation, in turn, sheds light on their validity.

Although  the Yudson approach  allows the exact determination the time evolution of any initial state  the hard problem of computing expectation values  of local operators in such state, as in any Bethe state,  still remains.  One approach, valid in the long time limit  is to replace integration over the parameters with their saddle point values. For attractive models, this requires separation of all free and bound states, thus one lose the compact form of the solution. On the other hand, one can make use of the form factor results. However, the form factor has apparent singularities when the parameters from the bra state and ket state overlaps. How the explicit results for the form factor survive for complex parameters is still an open question. Thus it is unclear if one can shift the integration contour to get rid of the poles in the form factor.

\section{Yudson Representation with Gaudin-Yang Eigenstates}
In this section, we will write down the explicit form of the Yudson representation with the Gaudin-Yang eigenstates in the coordinate basis. The integration contours will be specified for attractive and repulsive models respectively. Central theorem will be proved which shows that the Yudson representation is equivalent to the identity operator. 

Plugging in the result for $|\mu,k\rangle$ and $|\mu,k)$, it is easy to obtain the Yudson representation in real space as
\begin{align}\begin{split}
&\int_C dk\int_{C'} d\mu \langle y,\alpha|k,\mu\rangle(k,\mu|x,\beta\rangle\theta(x)\theta(y)\theta(\alpha)\theta(\beta)
 \\&= \sum_{P,R}(-1)^P e^{i\sum_i k_i(y_{\inv{P}i}-x_i)}\prod_{m<n}^M S(\mu_m-\mu_n)\prod_{m=1}^M J(\mu_m)\\
& \ \ \ \ \ \ \theta(x)\theta(y)\theta(\alpha)\theta(\beta)
 \end{split}\end{align}
Here $S(\mu-\nu)$ is defined in Eq.~\eqref{S_mu}, $J(\mu)$, which is an abbreviated notation of $J(\mu,k, P,\alpha,\beta)$ is defined as 
\begin{align}\label{eq1}
J(\mu)=&I(\mu,Pk,\alpha)I^*(\mu,k,\beta)\notag\\
=&\frac{-ic}{\mu-k_{P\alpha}+ic/2}\prod_{\substack{m<\alpha\\Pm\geq\beta}}\frac{\mu-k_{Pm}-ic/2}{\mu-k_{Pm}+ic/2}\notag\\
&\frac{ic}{\mu-k_{\beta}-ic/2}\prod_{\substack{n<\beta\\\inv{P}n\geq\alpha}}\frac{\mu-k_n+ic/2}{\mu-k_n-ic/2}
\end{align}
Here, we have taken into account the cancellation between $I(\mu, Pk, \alpha)$ and $I^*(\mu, k, \beta)$. However, to keep the expression compact, we have left the possible cancellation related to $k_{P\alpha}$ and $k_\beta$ unattended. We have only considered ordered positions($x$, $y$) and labels($\alpha$,$\beta$) as  any permutation on them does not create new states. 

\begin{figure}
\centering
\begin{minipage}{0.5\textwidth}
\centering
\subfloat[]
{
	\includegraphics[scale=0.3]{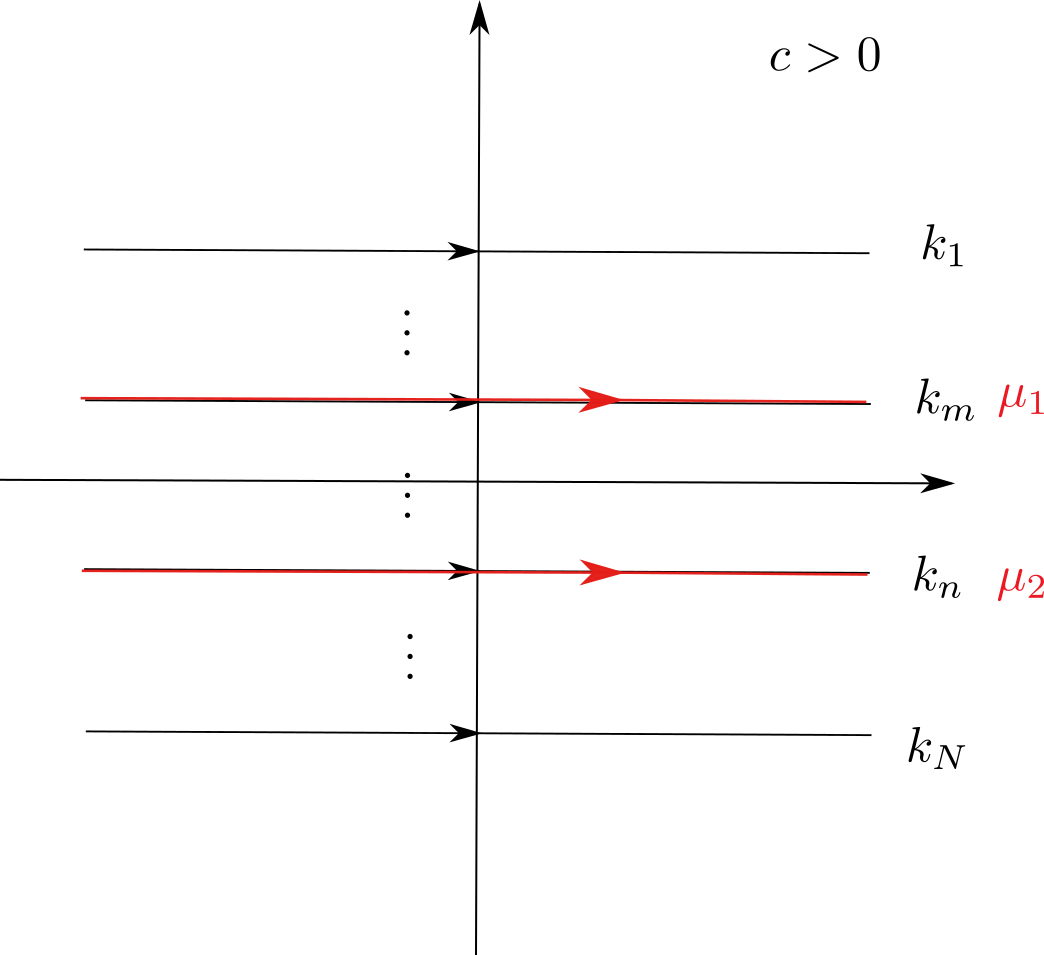}
	\label{fig:1}
}
\\
\subfloat[]
{
	\includegraphics[scale=0.3]{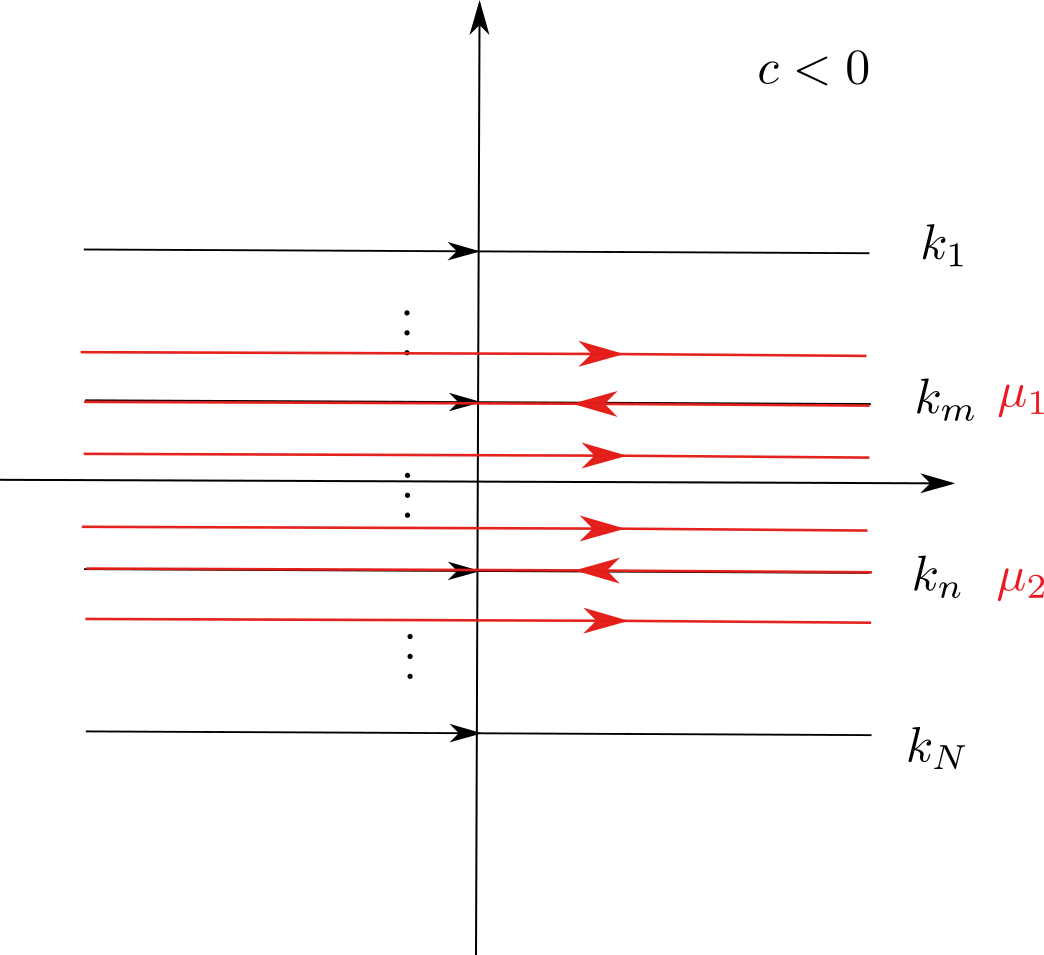}
	\label{fig:2}
}
\caption{Examples of integration contours in the Yudson representation for a system with $N-2$ majority fermions and two impurities. In this example, the $m$th and $n$th particles are the impurities counting from the left to right. Figure~\ref{fig:1} is for repulsive case and Figure~\ref{fig:2} is for attractive case.}
\label{GY_contour}
\end{minipage}
\end{figure}

As we have discussed in the previous chapter, the integration contour is of vital importance to the Yudson approach. For the Gaudin-Yang model, their choice depends on the initial condition. For the initial state $|x,\beta\rangle$, which means the N fermions locate at $x_1<\ldots<x_N$ and the down-spins correspond to label $\beta_1<\ldots<\beta_M$, the integral contours are chosen as follows. $k_1$ to $k_N$ are integrated along a horizontal direction. Their contours are separated by a distance greater than $2|c|$. The line of $k_1$ stays on the top and that of $k_N$ lies in the bottom. How $\mu$'s are integrated varies between repulsive and attractive models. For $c>0$, $\mu_m$ are integrated forward along the line of $k_{\beta_m}$. For $c<0$, $\mu_m$ are integrated backward along the contour of $k_{\beta_m}$ and forward along two lines that lie above and below that of $k_{\beta_m}$ with a separation greater than $|c|$ but still less than the distance between adjacent $k$. See Figure~\ref{GY_contour} as an example with $\beta_1=m$ and $\beta_2=n$. 
\subsection{Central Theorem}\label{sec:ct}
In this part, we will explain why such contour is chosen by proving the central theorem, which says that
\begin{align}\begin{split}
&\textit{Const}\int_C dk\int_C d\mu \langle y,\alpha|k,\mu\rangle(k,\mu|x,\beta\rangle\theta(x)\theta(y)\theta(\alpha)\theta(\beta)\\
&\ \ \ \ \ \ \ \ \ \ \ =\prod_i\delta(y_i-x_i)\prod_j\delta_{\alpha_j\beta_j}\theta(x)\theta(\alpha)
\end{split}\end{align}
The proof for repulsive cases and attractive cases is quite similar. We will illustrate the central theorem for repulsive model in detail and discuss how it can be applied to the attractive case. We will also start with the simplest situation where the system has a single fermion, which we will call an impurity, that is different from the other fermions, then move on to the multi-impurity scenario.
\subsubsection{Single Impurity Central Theorem for $c>0$}\label{sec: single}
When there is only one impurity, scattering among spin waves is absent. The central theorem simplifies to
\begin{align}\label{centralT_1}
\text{Const}&\int_C dk\int_{C'} d\mu \sum_P (-1)^Pe^{i\sum_i k_i(y_{\inv{P}i}-x_i)}J(\mu)\theta(y)\notag\\
&\theta(x)\theta(\alpha)\theta(\beta)=\prod_i\delta(y_i-x_i)\prod_j\delta_{\alpha_j\beta_j}
\end{align}
As $J(\mu)\sim \frac{1}{\mu^2}$ as $|\mu|\to\infty$, the integration contour of $\mu$ can be closed from above or below, either should yield the same result. We choose to close the contour in the upper half plane, then the $J$ integration transforms into sum of pole contributions. Denote $R(k_o+ic/2)$ as the residue of $J(\mu)$ at $\mu=k_o+ic/2$, then the $\mu$ integration simplifies as
\begin{align}\begin{split}
\int_C d\mu J(\mu)=2\pi i \sum_{\substack{o\leq \beta\\ \inv{P}o\geq\alpha}}R(k_o+ic/2)
\end{split}\end{align}
The $k$ integration in the central theorem~(\ref{centralT_1}) becomes
\begin{align}\begin{split}
2\pi i \int_C dk e^{i\sum_i k_i(y_{\inv{P}i}-x_i)}
\sum_{\substack{o\leq \beta\\ \inv{P}o\geq\alpha}}R(k_o+ic/2)
\end{split}\end{align}

Depending on the relation among $o$, $P\alpha$ and $\beta$, the limiting behaviour of $R$ as a function of $k_o$ can be categorized into three cases

\[
R(k_o+ic/2)=\left\{
\begin{array}{ll}
-ic+O(\frac{1}{k_o}) &\text{if }k_o=k_{P\alpha}=k_\beta\\
O(\frac{1}{k_o}) &\text{if }k_o=k_{P\alpha}\neq k_\beta\\
&\text{ or }k_o=k_\beta\neq k_{P\alpha}\\
O(\frac{1}{k_o^2}) &\text{In other cases}
\end{array}\right.
\]
Except the first term in the first line, all terms vanish as fast or faster than $1/k_o$ asymptotically. Using Jordan's lemma, the $k$ integration over these terms equals sum of pole residues above or below the integration contour, depending on the relation between elements of $P\vec{y}$ and $\vec{x}$. What we are going to show is that the total effect of these $O(\frac{1}{k^n})$ terms on the $k$ integration vanishes.

By partial fraction decomposition, the $O(1/k^n)$ terms can be turned into a sum of terms, with the summand takes the form
\begin{align}\begin{split}
E(k)\coloneqq E(k_1,\ldots, k_N)=\prod_{\{m\}}\frac{-ic}{k_o-k_{Pm}+ic}\prod_{\{n\}}\frac{ic}{k_o-k_n}
\end{split}\end{align} 
$m$ satisfies the condition $m\leq \alpha$ and $Pm\geq \beta$, $n$ satisfies the condition $n\leq\beta$ and$\invbig{P}n\geq\alpha$. Here both the product over $m$ and over $n$ can be empty product, but they cannot be empty at the same time. There are two types of poles in $E(k)$. One is of the form $k_o=k_n$. This pole is only apparent, but not real. As there is another term in $E(k_n)$ which cancels its contribution. To be explicit
\begin{align}\begin{split}
&\text{Res}\big(e^{i\sum_ik_{Pi}(y_{i}-x_{Pi})} R(k_o+ic/2), k_o=k_n\big)\\
&=-\text{Res}\big(e^{i\sum_ik_{Pi}(y_{i}-x_{Pi})} R(k_n+ic/2),k_n=k_o\big)
\end{split}\end{align}
\label{lltype}The other type of pole also exists in the Lieb-Liniger model. These Lieb-Liniger type of poles has the following character, all poles of $k_i$ above its integration contour locate at $k_j+ic$ with $i>j$ and $\invbig{P}i<\invbig{P}j$, all poles below locate at $k_m-ic$ with $i<m$ and $\invbig{P}i>\invbig{P}m$. Following the same argument as in~\cite{deepak1}, these poles do not contribute. Therefore, as part of the $k$ integrand, the $R(k_o)$ is equivalent to $-ic\delta_{P\alpha\beta}$ and the left hand side of Eq.~\eqref{centralT_1} becomes
\begin{align}\begin{split}
&(2\pi c) \times \text{Const}\int_C dk \sum_P(-1)^P e^{i\sum_i k_i(y_{\inv{P}i}-x_i)}\delta_{P\alpha\beta}\\
&\ \ \ \ \ \ \theta(x)\theta(y)\theta(\alpha)\theta(\beta)\\
=&(2\pi)^{N+1}c \times \text{Const} \prod_i \delta(y_i-x_i)\delta_{\alpha\beta}\theta(x)\theta(\alpha)
\end{split}\end{align}
This completes the proof of central theorem for system with single impurity and fixes  the normalization constant as $1/((2\pi)^{N+1}c$.

\subsubsection{Multi-impurity Central Theorem for $c>0$}\label{sec:mltyCT}
The central theorem for multiple down spins states,
\begin{align}\label{centralT2}
&\int_C dk \int_{C'} d\mu \sum_{P,R}(-1)^P e^{i\sum_i k_i(y_{\inv{P}i}-x_i)}\prod_{m<n}S(\mu_m-\mu_n)\notag\\
&\prod_{m=1}^M J(\mu_m)\theta(x)\theta(y)\theta(\alpha)\theta(\beta)=\prod_i\delta(y_i-x_i)\prod_m\delta_{\beta_m\alpha_m}
\end{align}
The proof is similar to that of the single impurity case. First, one carry out the integration over $\mu_1,\ldots,\mu_M$ respectively by closing each integral contour in the upper half plane. As we will see, this leads to an expression with only Lieb-Liniger type of pole for the $k$ integration. Then one performs the $k$ integration and makes the same argument as we do in the previous case.

First, complete the $\mu_1$ integration. Since the pole of $S(\mu_1,\mu_m)$($m>1$) is at $\mu_1=\mu_m+ic$, which is below the $\mu_1$ contour, this pole is not included. Thus, the integral becomes sum of residues at the same set of poles as that with $J(\mu_1)$ alone. In the meanwhile, the factor $S(\mu_1,\mu_m)$ becomes $(\mu_m-k_o-ic/2-ic\Sgn(\alpha_1-\alpha_m))/(\mu_m-k_o+ic/2)$. Therefore, we have
\begin{align}\begin{split}
&\int d\mu_1 J(\mu_1)\prod_{m>1}S(\mu_1-\mu_m)=(2\pi i)\sum_{o}R(k_o+ic/2)\\ &\ \ \ \ \ \ \ \ \prod_m \frac{\mu_m-k_o-ic/2-ic\Sgn(\alpha_1-\alpha_m)}{\mu_m-k_o+ic/2}
\end{split}\end{align}
which summed over terms of all poles lying above the $\mu_1$ integration contour, i.e. $o\leq\beta_1$ and $\invbig{P}o\geq\alpha_1$. 

Then, we perform the integration over $\mu_2$. Among the S-matrices, only $S(\mu_1-\mu_2)$ contributes a pole that get enclosed by the contour, which is located at $\mu_m=k_o+ic/2$ if $\alpha_1>\alpha_2$. Combined with the condition on $o$, $\alpha_1$, $\alpha_2$ and $\beta_1$, $\beta_2$, we have $\invbig{P}o\geq\alpha_1>\alpha_2$ and $o\leq\beta_1<\beta_2$. Thus $\frac{\mu_2-k_o+ic/2}{\mu_2-k_o-ic/2}$ must be a factor of $J(\mu_2)$. It cancels the denominator of $S(k_o+ic/2-\mu_2)$. Therefore, the poles above the $\mu_2$ contour in $J(\mu_2)S(k_o-\mu_2+ic/2)\prod_{m>2}S(\mu_2-\mu_m)$ is the same as that in $J(\mu_2)$. And
\begin{align}\begin{split}
&\int d\mu_2 J(\mu_2)S(k_0+ic/2-\mu_2)\prod_{n>2}(\mu_2-\mu_n)
=(2\pi i)\\& \sum_{p}R^{(1)}(k_p+ic/2)\prod_{n>2}  \frac{\mu_n-k_p-ic/2-ic\Sgn(\alpha_2-\alpha_n)}{\mu_n-k_n+ic/2}
\end{split}\end{align}
Here $R^{(1)}(k_p+ic/2)=R(k_p+ic/2)(k_p-k_0+ic+ic\theta(\alpha_1-\alpha_2))/(k_p-k_o+ic)$. Since the denominator $k_p-k_o+ic$ is cancelled by a numerator in $R(k_p+ic/2)$, $R^{(1)}$ has the same denominator and asymptotic behaviour as $R$.

The argument can be generalized to any $\mu$ integration. After all $\mu$ integration is done, the result becomes
\begin{align}\begin{split}
&\int d\mu \prod_{m<n} S(\mu_m-\mu_n) \prod_m J(\mu_m) =(2\pi i)^M \sum_{o_1,\ldots, o_M} R(k_{o_1}\\&\ \ \ \ \ \ +ic/2) R^{(1)}(k_{o_2}+ic/2)\ldots R^{(M-1)}(k_{o_m}+ic/2)
\end{split}\end{align}

$R^{(n)}$ are generated by the $\mu_{n-1}$ integration, it has the same pole structure and asymptotic limit as $R$. As shown for the single impurity case, all poles of $R$ are of the Lieb-Linger type that satisfies the condition discussed on page~\pageref{lltype}, so is it for $R^{(n)}$ and $R$ product.

Thus, the $k$ integration in the central theorem with multiple down spins is essentially the same as that for the single impurity case, with $R$ replaced by the $R$ product. Now, the product of $R$ contribute a $(-ic)^M$ if $k_{o_m}=k_{P\alpha_m}=k_{\beta_m}$ for $m=1,\ldots,M$. In other cases, it is of the order $O(1/|k|)$ or $o(1/|k|)$ with no pole contributes to the integral. Therefore, the left hand side of Eq.~\eqref{centralT2} becomes
\begin{align}\begin{split}
&(2\pi c)^M \times \text{Const}\int_C dk \sum_P(-1)^P e^{i\sum_i k_i(y_{\inv{P}i}-x_i)}\\
&\ \ \ \ \ \ \prod_m\delta_{P\alpha_m\beta_m}\theta(x)\theta(y)\theta(\alpha)\theta(\beta)\\
=&(2\pi)^{N+M}c^M \times \text{Const} \prod_i \delta(y_i-x_i)\delta_{\alpha\beta}\theta(x)\theta(\alpha)
\end{split}\end{align}
This completes the proof for multi-impurity central theorem and fix the normalization constant as $1/((2\pi)^{N+M}c^M)$.
\subsubsection{Central Theorem for $c<0$}
\begin{figure}
	\includegraphics[scale=0.28]{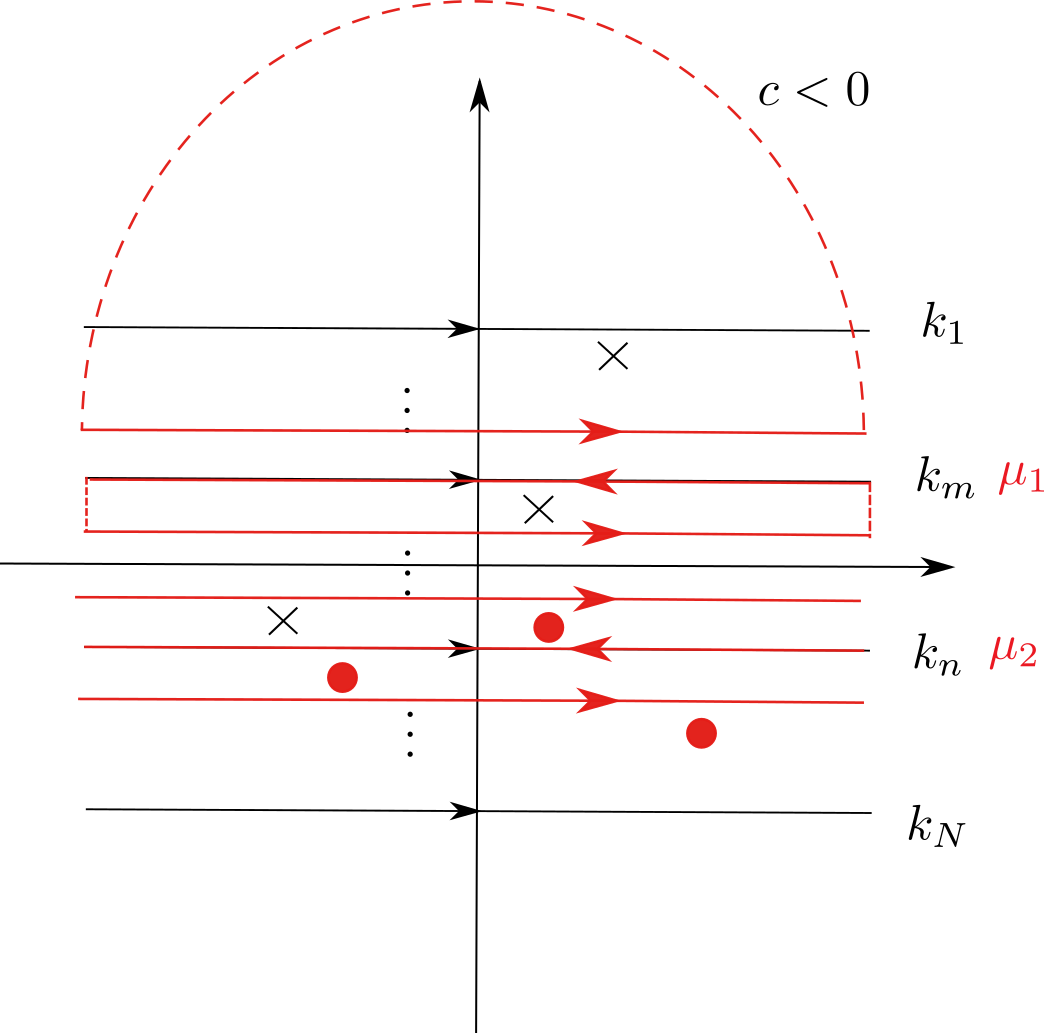}
	\caption{Distribution of poles for $\mu$ integration in attractive model. Crosses represent poles from $J(\mu)$, dots denote poles from $S(\mu_m-\mu_n)$}
	\label{attractivePole}
\end{figure}

The central theorem of the Yudson representation for the attractive Gaudin-Yang model is essentially the same as that for the repulsive case. Although the relative position between the poles and the contours are different, the choice of the contours guarantees that the same set of poles contribute to the integration as in the repulsive case. To see this, consider the $\mu$ integration first. While in the repulsive case, the poles from $J(\mu)$ that contribute after closing the contour from above are $k_o+ic/2$ with $o\leq\beta$ and $\invbig{P}o\geq\alpha$. Now with $c<0$, the pole at $k_\beta+ic/2$ are below the integration contour of $\mu$. But with the three line contour, this pole will also be enclosed by closing the contour counterclockwise, see Figure~\ref{attractivePole}.
The other poles with $o<\beta$ remains above the contour of $\mu$. The scattering matrices leads to poles at $\mu_m=\mu_i+ic$ for $i>m$ and $\mu_m=\mu_j-ic$ for $j<m$. For repulsive case, the first type of poles stay below the integration contour while the second lie above. This still holds for attractive cases. Therefore, the $\mu$ integration results in the same expressions as in the previous case, with only Lieb-Liniger type of pole, i.e. $k_i-k_j+ic=0$ with $i<j$ and $\invbig{P}i>\invbig{P}j$. In the repulsive case, the pole of $k_i=k_j-ic$ is below the line of $k_i$ integration, this is still the case for attractive scenario as contour of $k_j$ is below that of $k_i$ by a distance greater than $2|c|$. Therefore, the argument for repulsive cases also applies to attractive cases. Thus we will not repeat the proof for $c<0$ and central theorem holds for any type interaction.

\section{Bound States}\label{sec:bs}
 The completeness of the representation is proven directly via contour integrals and does not requires a previous knowledge of the eigenstates. This feature, as mentioned earlier allows us to determine the actual string structure of the system in the infinite volume limit, as the contour includes both free state - corresponding to real momenta  $k$'s  and  bound states  corresponding to complex  momenta. In this section, we will separate out these bound states. The aim of this is twofold. First, it tests the validity of the String hypothesis by enumerating every bound states in the complete basis. Second, since these bound states have different energy and spread out at various speed, their contributions separate apart asymptotically in real space and we need such separation in calculations for asymptotic behaviours.

Before we carry out  the separation, we will define five kinds of string solutions
\begin{enumerate} 
\item $k-\mu$ string of length $n$.\\
 $n$ $k$'s forming a string with $n-1$ $\mu$'s, with $k=\bar{k}+ic(j-n/2-1/2)$ for $j=1,\ldots, n$ and $\mu=\bar{k}+ic(j-n/2)$ for $j=1,\ldots,n-1$.
\item $\mu-k$ string of length $n$.\\
 $n$ $\mu$'s form a string with $n-1$ $k$'s, with $k=\bar{k}+ic(j-n/2)$ for $j=1,\ldots,n-1$ and $\mu=\bar{k}+ic(j-n/2-1/2)$ for $j=1,\ldots, n$.
 \item $\mu$ string of length n\\
 $n$ $\mu$'s form a string without $k$, with $\mu=\bar{\mu}+ic(n/2+1/2-j)$ for $j=1,\ldots, n$.
 \item $k$ string of length n\\
 $n$ $k$s form a string without $\mu$, with $k=\bar{k}+ic(n/2+1/2-j)$ for $j=1,\ldots, n$.
 \item $k-\mu$ pair.\\
One real $k$  pairs with one  $\mu$, either  $\mu=k\pm ic/2$    or   $\mu=k - ic/2$, depending on the representation (See below).  These pairs have no physical consequences.
\end{enumerate}
\begin{figure}
	\centering
	\includegraphics[width=0.45\textwidth]{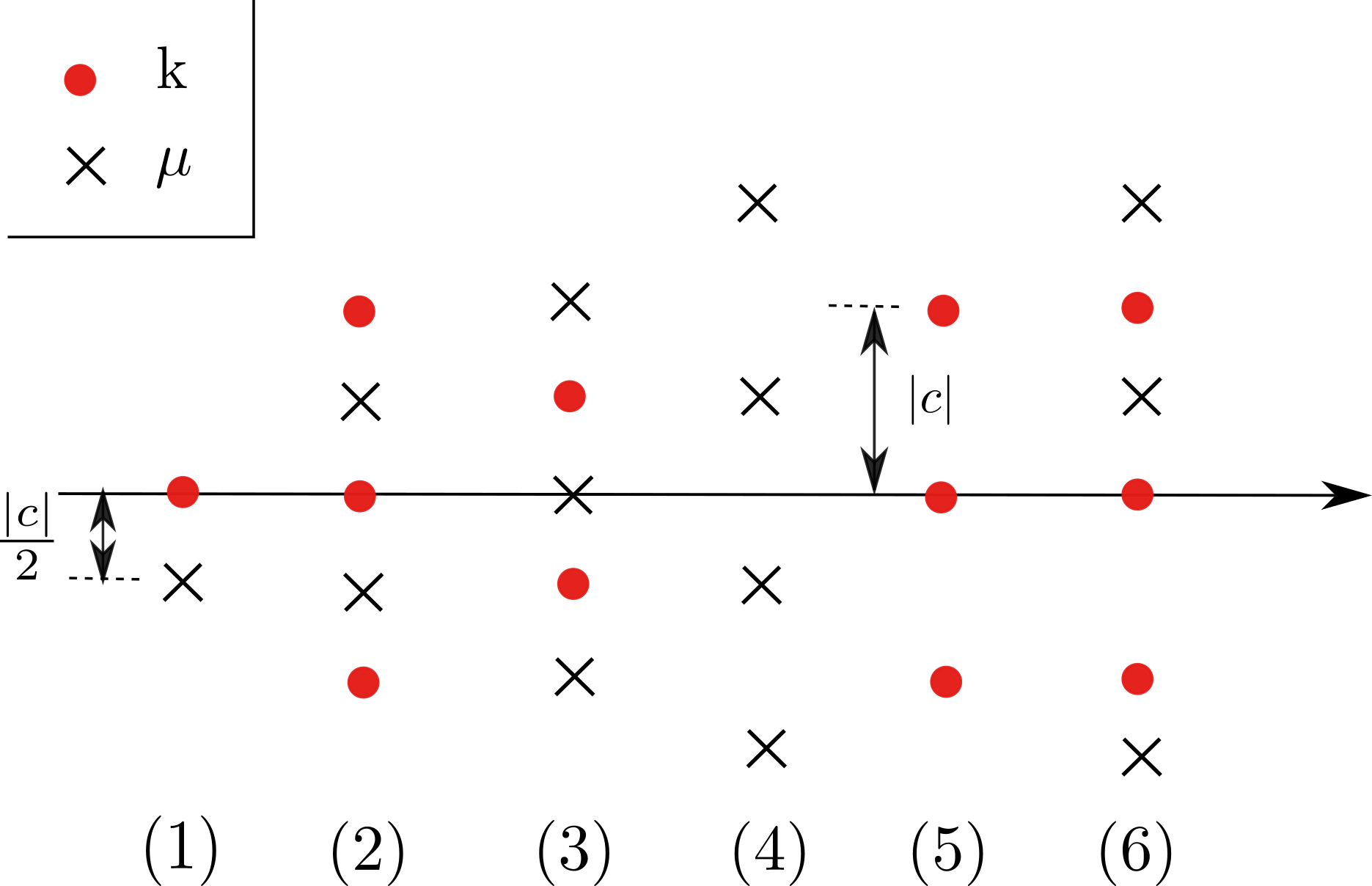}
\caption{Illustration of various string solutions. (1) k-$\mu$ pair. (2)k-$\mu$ string of length 3. (3)$\mu$-k string of length 3. (4)$\mu$ string of length 4. (5)k string of length 3. (6)Composite of 3 k-$\mu$ pairs and one $k$ string of length 3 (only exists in bosonic Gaudin-Yang model, see chapter~\ref{sec:BGY_boundstates}.)}
\label{fig:stringtype}
\end{figure}
Figure~\ref{fig:stringtype} shows an illustration of each type of string solutions. In the String hypothesis, Takahashi predicted the existence of $\mu$ string of any length and $k-\mu$ string of length 2. In this section, we will see strings of the first two  as well as the last types emerging by shifting contours. No strings of $\mu$ or $k$ alone exist. $k-\mu$ pair is a basic building block of string solutions, but it does not have any physical significance in real space. $\mu-k$ strings of length $n+1$ is treated as $k-\mu$ string of length $n$ accompanied by two $k-\mu$ pairs. The two strings are physically equivalent. Moreover, we will see that although $\mu-k$ strings of length 2 describes bound state between two down spins, it can be decomposed into sum of free states. For $c<0$, we will also see $k-\mu$ string of length $n>2$, which represents a cluster of $n$ particles bounded together by $n-1$ spin waves. 

In order to perform the separation, we first perform the $\mu$ integration. This will result in a constant term plus a sum over terms with Lieb-Liniger type of poles, i.e. $k_i-k_j+ic=0$ for $i<j$ and $\invbig{P}i>\invbig{P}j$. Then one shift all $k$ contours to the real axis. For $c>0$, the pole of $k_i=k_j-ic$ lies below both $k_i$ and $k_j$ contours.  Thus, no pole is caught when one shift contours to the real axis. For this reason, the only type of string exists, after separating different states apart, is the one with a real $k_i$ ($i\leq\beta$ and $\invbig{P}i\geq\alpha$) and a complex $\mu$ at $k_i+ic/2$ (string type 1). However, such string depends on how we close the $\mu$ contours. If we close both contours from below, we are left with a different string configuration with $\mu=k_j-ic/2$ for $j\geq\beta$ and $\invbig{P}j\leq\alpha$. Physically, it should not matter how one closes the contour, as mathematically it leads to the same result. Indeed, they are not related to bound state in real space.

However, we also noticed the appearance or disappearance of a $\mu$-k string of length 2 which depends on how one closes the contour. 
For the purpose of illustration, we will consider a specific scenario, with $\alpha_1=3,\alpha_2=2$, $\beta_1=2,\beta_2=4$, $P=P_{15}$. Under this condition, the integrand of $\mu$ and $k$ becomes
\begin{align}\begin{split}
S(\mu_1-&\mu_2)\prod_{i=1}^2 J(\mu_i,Pk,\alpha_i)J^*(\mu_i,k,\beta_i)=\frac{\mu_1-\mu_2+ic}{\mu_1-\mu_2-ic}\\
&
\times\frac{\mu_1-k_1+ic/2}{\mu_1-k_1-ic/2}\frac{ic}{\mu_1-k_2+ic/2}\frac{-ic}{\mu_1-k_3+ic/2}\\&
\times\frac{\mu_1-k_5-ic/2}{\mu_1-k_5+ic/2}\frac{\mu_2-k_1+ic/2}{\mu_2-k_1-ic/2}\frac{-ic}{\mu_2-k_2-ic/2}\\
&\times\frac{\mu_2-k_3+ic/2}{\mu_2-k_3-ic/2}\frac{ic}{\mu_2-k_4-ic/2}\frac{\mu_2-k_5-ic/2}{\mu_2-k_5+ic/2}\\
\ 
\end{split}
\end{align}
If one closes the contour of $\mu_1$ and $\mu_2$ from the same direction, there is no $\mu-\mu$ string. But if one closes $\mu_1$ from below and $\mu_2$ from above, we see four terms corresponding to $\mu_1=\mu_2+ic=k_i+3ic/2$ for $i=1,2,3,4$. What is strange about these solutions is that it leads to a result whose denominator has poles at $k_i-k_j-2ic$. This is a new type of pole. But the fact is, if we sum up the contribution from these strings, these new poles are canceled by the numerator of the sum. This means, the sum of the $\mu$ strings is equivalent to a collection of states without strings. Physically, that means the bound state between two down spins can be decomposed into sum of free states. As the energy does not depend on $\mu$'s, the bound states keep a coherence phase with all other states. Thus, it is impossible to tell whether the decrease of the wavefunction as the two impurities separate is due to bound states or simply destructive interference among free states. That is to say, the $\mu$ string does not play an important role in real space. As there is no $k$ strings when $c>0$, there is no bound states in such system. 

\begin{figure}[h]
\centering
	\begin{minipage}{0.45\textwidth}
	\centering
	\subfloat
	{
		\includegraphics[width=3.8cm]{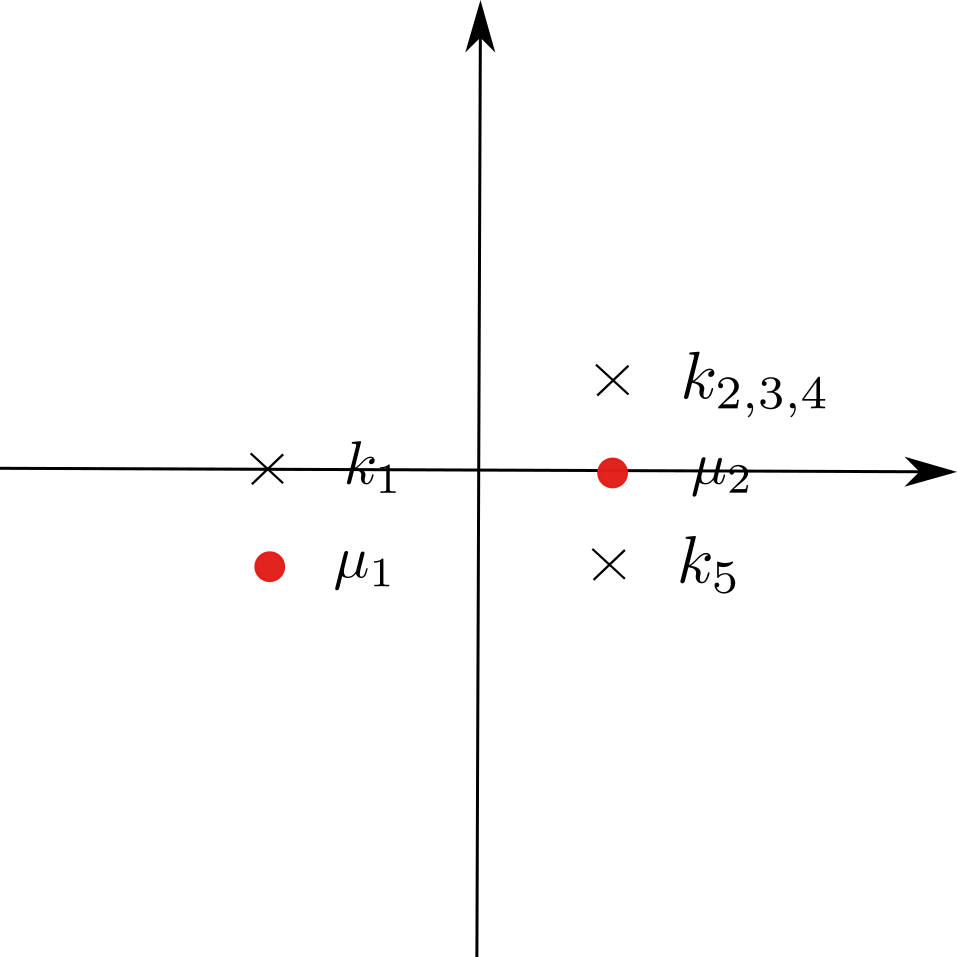}
		\label{f-a}
	}
	\subfloat
	{
		\includegraphics[width=3.8cm]{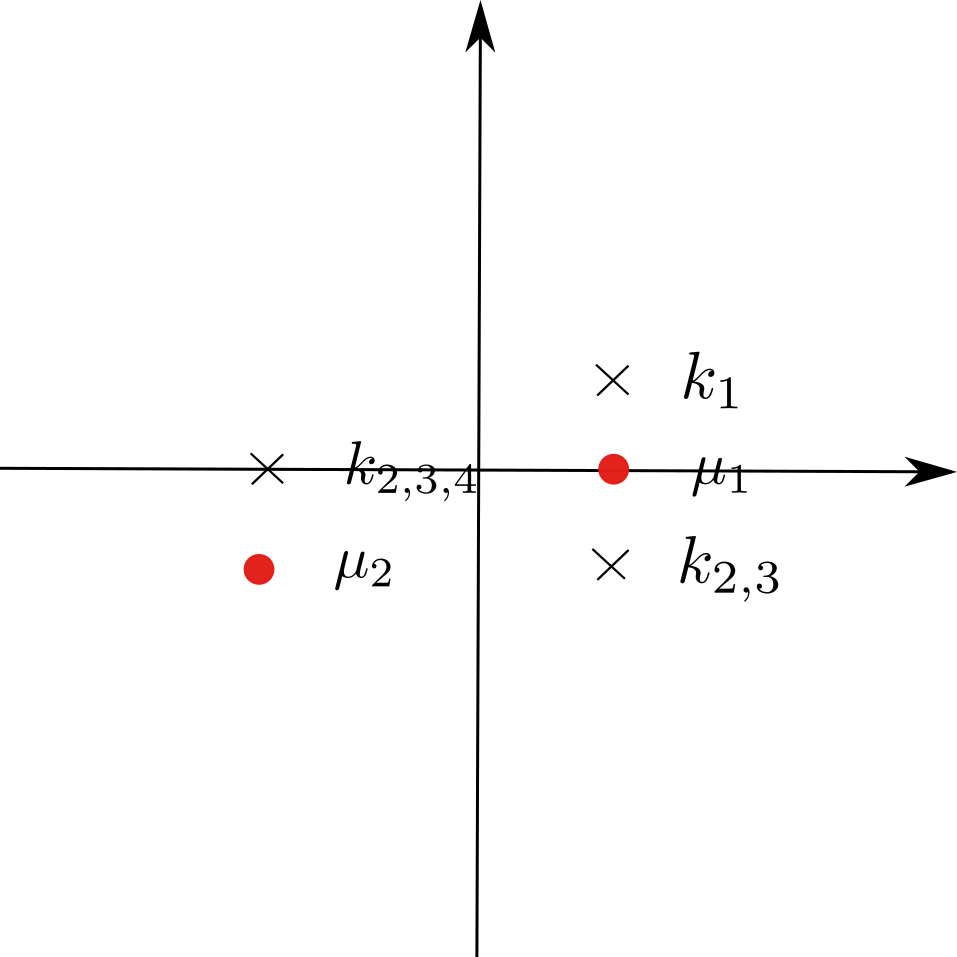}
		\label{f-b}
	}
	\\
	\subfloat
	{
		\includegraphics[width=3.8cm]{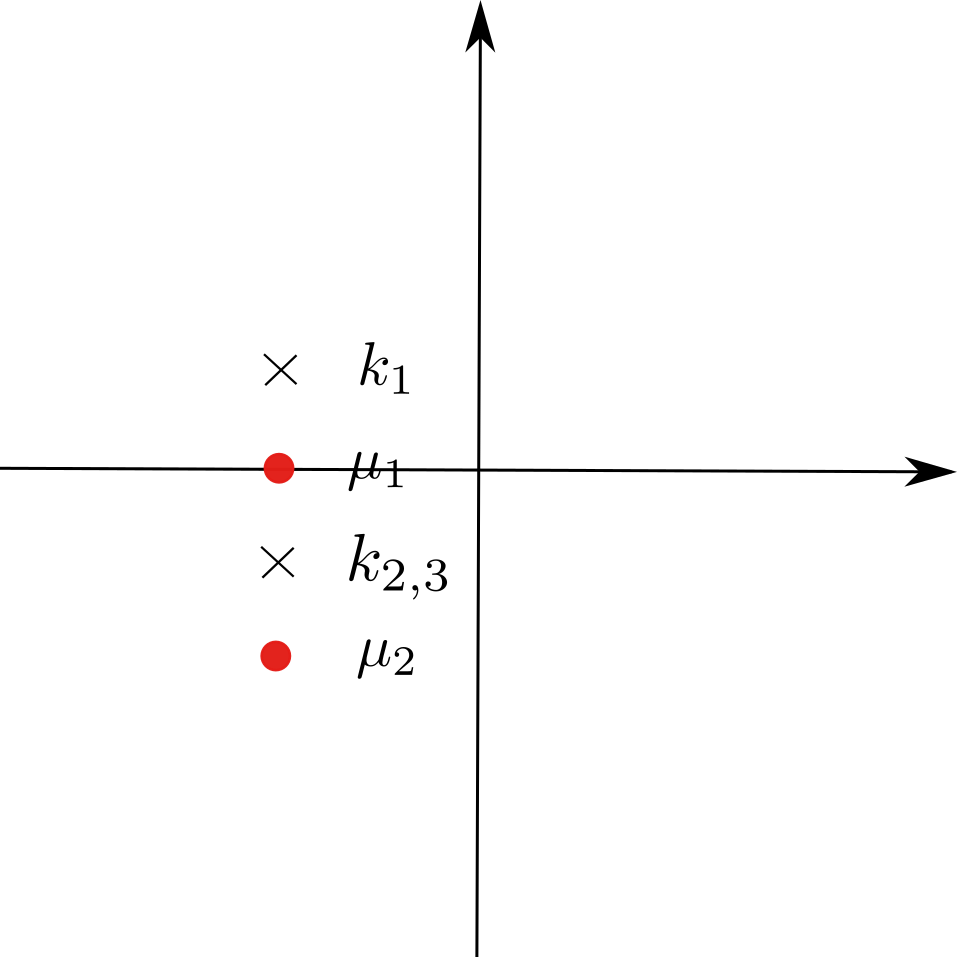}
		\label{f-c}
	}
	\subfloat
	{
		\includegraphics[width=3.8cm]{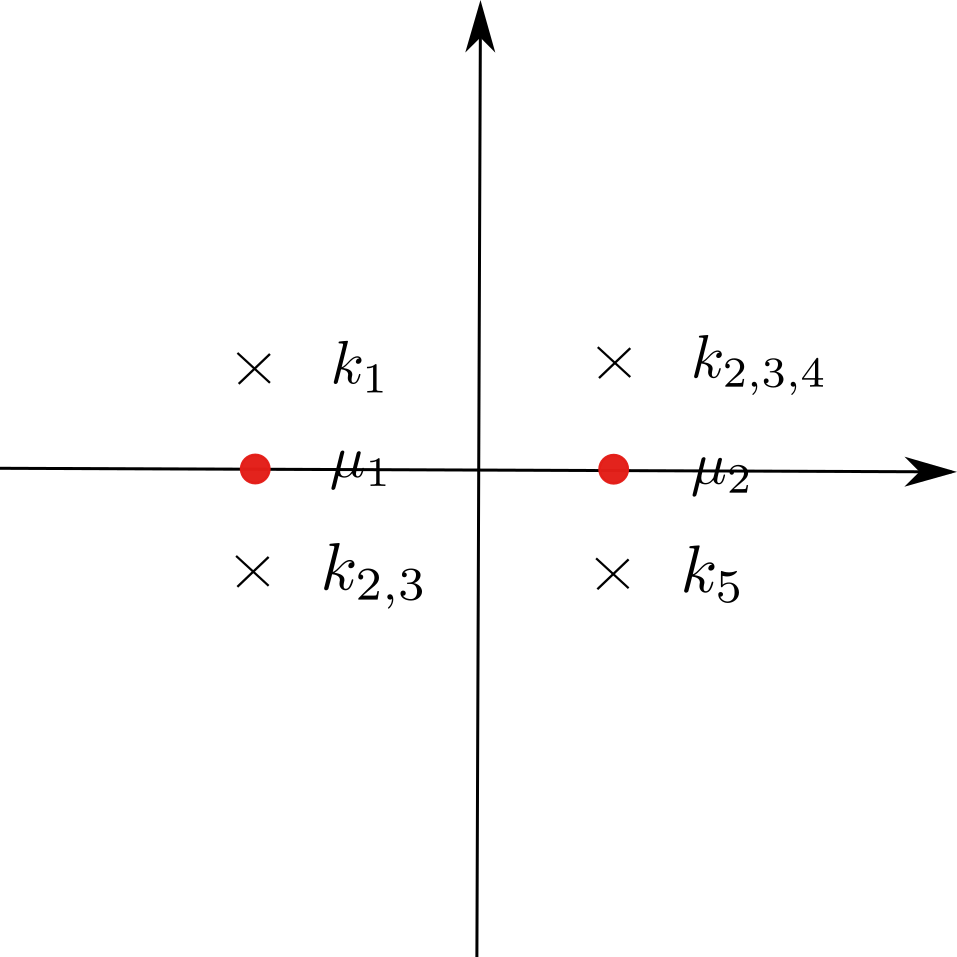}
		\label{f-d}
	}
	\\
	\subfloat
	{
		\includegraphics[width=3.8cm]{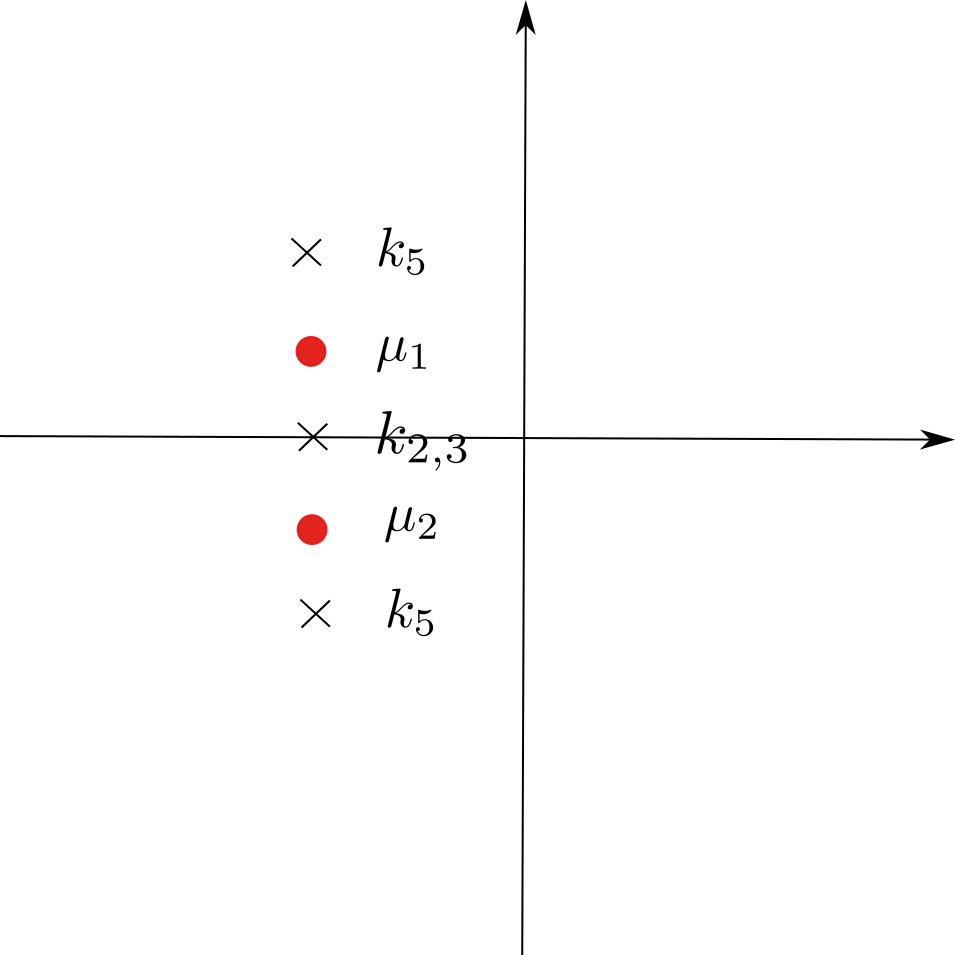}
		\label{f-e}
	}
	\caption{String solutions in state $\beta_1=2,\beta_2=4,\alpha_1=3,\alpha_2=2,P=P_{15}$ after closing $\mu$ contours in the upper half plane. Multiple subindex of the $k$ represents a $k$ with one of the subscript as long as it does not coincide with any other $k$ subscripts in the plot. The plot intends to show the relative position among parameters in each strings or pairs. The real part of them is to be integrated over. Any $k$'s that are not shown explicitly are assumed to be integrated along the real line.}
	\label{fig:st}
	\end{minipage}
\end{figure}

When $c<0$, the pole of $k_i$ at $k_j-ic$ is between the contours of $k_i$ and $k_j$. When shifting both contour to the real axis, the original integral splits into two terms. One results from the residue, the other one is integrated along the real axis. The former describes a bound state and the other is a free state, see~\cite{deepak1}. As the appearance of this pole originates from the residue at $\mu_m=k_i+ic/2$, such pole contribution yields bonds among $k_i$, $\mu_m$ and $k_j$. Since no two $k$'s can take the same value, each $\mu$ connects at most two $k$'s. However when different $\mu$'s relate to a same $k$, it will snap two $k-\mu$ strings together to form a longer string. Such $k-\mu$ string solutions of length greater than 2 are not included in the String hypothesis, as it involves ambiguity of $0/0$ in the Bethe equations Eq.~\eqref{bethe1} and \eqref{bethe2}. However, they do exist in the time evolution of the system. Figure~\ref{fig:st} shows a complete set of string solutions in the example considered here.

In order to interpret these string solutions, we keep track of the particles by their quasimomentum $k$'s. In the above example, we have $P=P_{15}$, thus particle $k_1$ switches position with particle $k_5$. In the final state, particle $k_{Pi}$ becomes the $i$th particle in the sequence. $\invbig{P}i$ is the location of the particle $k_i$ in the state.

When $P\alpha=\beta$, i.e. the down spins stays with the original particles, the integral does not vanish under any circumstances. However when $P\alpha<\beta$, i.e. the down spin transfers to a prior particle labeled by $k_{P\alpha}$, the integral will vanish unless there exists a $k_{Pi}$ such that $i<\alpha$ and $Pi\geq\beta$. This indicates that a particle may acquire the down spin in two ways. Either the impurity itself crosses the target particle or the particle after $\beta$ first passes the impurity and then pass the target particle. 

Besides transferring down spins, the interaction can also lead to bound states. In order to form a bound state between particle $i$ and $j$, the following condition must be satisfied $i\leq\beta\leq j$ and $\invbig{P}i\geq\alpha\geq\invbig{P}j$. Physically, this means one of the following situations depicted in table 1. When a bound state involves an up spin and a down spin, it is a singlet which is related to Cooper pair in BCS and FFLO state. When the bound state includes two up spins, it is a triplet and describes a bound state in a normal state.

\begin{table}[h!]
\centering
\begin{tabular}{c}
\includegraphics[width=0.9\linewidth]{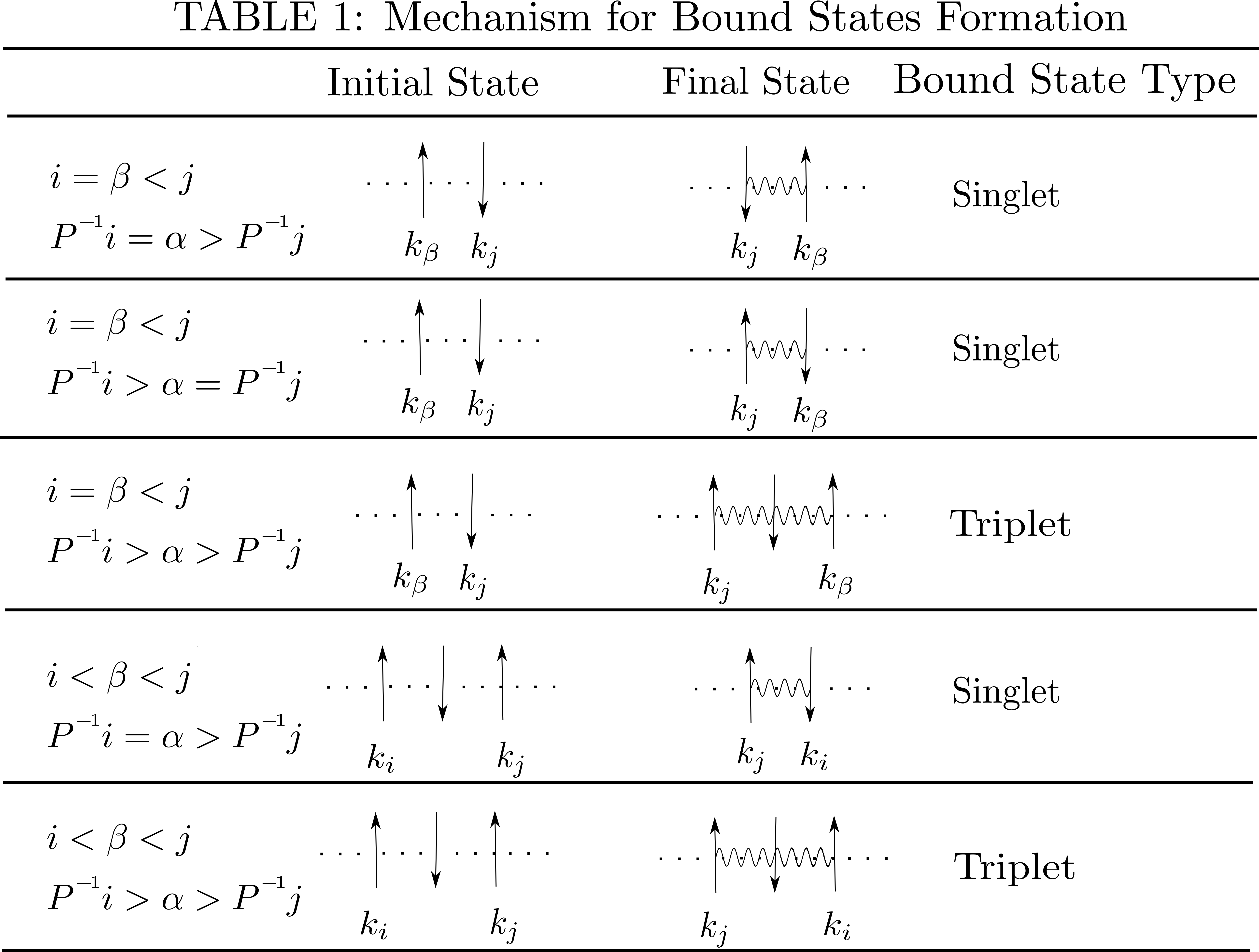}   
\end{tabular}
\label{fig:boundstate}
\end{table}

As shown in~\cite{guan2013fermi}, in an unpolarized system, the singlet bound state leads to Cooper pairs in the weak interaction regime, and it relates to the bound state with bosonic nature that results in BEC with strong attraction. When the spin is unbalanced, the string solution characterizes FFLO states. However, the bound state that is described by the string solution, is not a cooper pair in BCS or FFLO state by itself. The center of mass momentum of the former is determined by the spin rapidity, while that of the latter equals zero or the difference between the Fermi wave vectors. In fact, the Cooper pair is a \textit{dressed} version of the bound state described by the string solution. That is to say, changing one string solution will affect the position of other strings as determined by the Bethe equation. And the cooper pair describes the collective behavior of the whole system. When the spin is unpolarized, breaking a string does not change the total momentum as the other string will shift their center accordingly to minimize the energy and remain symmetric on the momentum axis. The situation is different when there are unpaired fermions, which interact differently with the strings. Thus breaking a string will change the total momentum, making the momentum of the \textit{dressed} bound state nonzero. Here we only provide an intuitive picture between the string solutions and the Cooper pair. The explicit relation between them is still elusive. But one can tell that the existence of the singlet solution is a necessary condition to the emergence of the FFLO state.

Here we want to remark that the situation is quite different in a quench dynamic where the system no longer seeks the lowest energy state. Instead, each string solution evolves independently with the weight determined by the overlap. Due to the abundance of excitation in the nonequilibrium dynamics, the FFLO \textit{quasiparticle} picture does not hold. Indeed, as shown in~\cite{bolech2012long}, the pair correlation loses the FFLO signature quickly after turning off the longitudinal confinement potential.

Aside from bound states between two particles, more particles can bind together via more spin wave modes. In the example in Figure~\ref{f-e},  $k_1$, $k_2$ and $k_5$ forms a bound state with the help of $\mu_1$ and $\mu_2$. In order to form such a bound state, one needs to reverse the order of particle 1,2,4 and 5. Though particle 4 does not show up in the bound state, the spacial motion of it relative to the particles in the bound state is crucial to the formation of it. Moreover, as particle 4 is sandwiched between particle 2 and particle 5, the wavefunction will also decrease exponentially with the separation between particle 4 and 2 as well as particle 4 and 5. However, such binds are different to that among $k_1$, $k_2$ and $k_5$. The wavefunction of this bound state consists of the factor $\exp(ik(y_{\inv{P}1}-x_1+y_{\inv{P}2}-x_2+y_{\inv{P}5}-x_5)+c/2(y_{\inv{P}1}-x_1-y_{\inv{P}5}+x_5))$ $\exp(ik_4(y_{\inv{P}4}-x_4))$. Therefore, although particle 4 is bounded with particle 1,2 and 5, it does not move coherently with them and its quasimomentum is different from that of the center of mass of the bound state. Note, such bound state cannot be decomposed into two bound states related to shorter strings, which should take the form $\exp((ik(y_{\inv{P}1}-x_1+2y_{\inv{P}2}-2x_2+iy_{\inv{P}5}-x_5)+c/2(y_{\inv{P}1}-x_1-y_{\inv{P}5}-x_5))$. Thus, in the String hypothesis, one should include $k-\mu$ strings of all lengths.

%and only include $k-\mu$ strings in the complete basis.

\section{Time Evolution}
In this section, we will apply the Yudson approach to study the quench dynamics. As shown in the previous sections, the Yudson representation provides us with an expansion of a state into components that are eigenstates of the Hamiltonian. Each component evolves with the factor $e^{-iEt}$ with $E$ being the eigenenergy of the basis state. Theoretically, this representation solves the time evolution of any initial state. However, due to the structure of the Bethe eigenstate, we will consider only the case where particles are well separated. In this work, we will study the initial state where particles are described by Gaussian wavepacket of width $\sigma$ and mean $x_{i0}$, with $x_{(i+1)}-x_{i}=a>3\sigma$. This problem has been studies in finite systems with the periodic boundary condition, where oscillatory behavior is observed as a sequence of backward scattering at the boundary \cite{robinson2016exact,robinson2016motion}. Here, we will instead consider an open system. The initial state $|\phi_0\rangle$ can be described as 
\begin{align}\begin{split}
|\phi_0\rangle
=\frac{1}{(\pi\sigma^2)^{N/4}}\int dx' e^{-\sum_i^N\frac{({x'}_i-x_{i})^2}{2\sigma^2}}|x',\beta\rangle
\end{split}\end{align}
Physically, such states are easily prepared with optical potential. Theoretically, the state saves us from all Heaviside step functions of coordinate $x$'s, which leads to huge complications in calculating observables. Moreover, the width $\sigma$ serves as a convergent factor for the $k$ integration, as we will see in the following calculation. With the Yudson representation, the time evolved state can be written as
\begin{align}\begin{split}
&|\phi(t)\rangle\\
=&\frac{(4\pi\sigma^2)^{\frac{N}{4}}}{(2\pi)^{M+N}c^M}\int dy \sum_\alpha \int_C dk \int_{C'} d\mu \sum_{P,R}(-1)^P e^{-i \sum_i k_i^2t}\\
&e^{-\sum_i k_i^2\sigma^2/2+i\sum_i k_i(y_{\inv{P}i}-x_{i})}\prod_{m<n}S(\mu_m-\mu_n) \prod_m J(\mu_m)\\
&\theta(y)\theta(\alpha)|y,\alpha\rangle+O(e^{-\frac{a^2}{4\sigma^2}})
\end{split}\end{align}

\begin{align}
\begin{split}
S(\mu_m-\mu_n)=\frac{\mu_m-\mu_n+ic\Sgn(\alpha_{
\inv{R}m}-\alpha_{\inv{R}n})}{\mu_m-\mu_n-ic}
\end{split}
\end{align}

\begin{align}
\begin{split}
J(\mu)
=&\frac{-ic}{\mu-k_{P\alpha}+ic/2}\prod_{\substack{m<\alpha\\Pm\geq\beta}}\frac{\mu-k_{Pm}-ic/2}{\mu-k_{Pm}+ic/2}\\
&\frac{ic}{\mu-k_{\beta}-ic/2}\prod_{\substack{n<\beta\\\inv{P}n\geq\alpha}}\frac{\mu-k_n+ic/2}{\mu-k_n-ic/2}
\end{split}
\end{align}
\subsection{Dynamics of Two Distinct Fermions}
We will first study the simplest case, a system with one down spin at $x_{10}$ and one up spin at $x_{20}=x_{10}+a$. This is the only scenarios where a closed form of the wavefunction can be obtained, which equals
\begin{align}\label{2w}
&f_{\uparrow,\downarrow}(y_1,y_2)\notag\\
=&\frac{\sigma}{2\sqrt{\pi}i (t+\sigma^2/2i)}\big(e^{\frac{i(y_2-x_1)^2}{4(t+\sigma^2/2i)}+\frac{i(y_1-x_2)^2}{4(t+\sigma^2/2i)}}(1\mp \
c/2\notag\\
&(1+i)\theta(y_2-y_1)\sqrt{\pi(t+\sigma^2/2i)}\erfc(\pm \alpha_1)e^{\alpha_1^2}\big)\notag\\
& \mp e^{\frac{i(y_1-x_1)^2}{4(t+\sigma^2/2i)}+\frac{i(y_2-x_2)^2}{4(t+\sigma^2/2i)}}c/2(1+i)\theta(y1-y2)\sqrt{\pi}\notag\\
&\sqrt{(t+\sigma^2/2i)}\erfc(\pm\alpha_2)e^{\alpha_2^2}
\end{align}
\begin{align}\begin{split}
\alpha_1= \frac{(1-i)(y_2-x_1-y_1+x_2+2ic(t+\sigma^2/2i)}{4\sqrt{(t+\sigma^2/2i)}}
\end{split}\end{align}
\begin{align}\begin{split}
\alpha_2= \frac{(1-i)(y_1-x_1-y_2+x_2+2ic(t+\sigma^2/2i)}{4\sqrt{(t+\sigma^2/2i)}}
\end{split}\end{align}
The $\pm$ sign results from the ambiguity of pulling out the $i$ from the square root. Depending on the phase of $\alpha_{1,2}$, the signs are chosen as follows
\begin{enumerate}
\item When $c>0$, $|\text{ph}(\alpha_{1,2})|\in(0,\pi/4)$. The upper sign is chosen.
\item When $c<0$ and $|c|>2a/\sigma^2$, $|\text{ph}(\alpha_{1,2})|\in (\frac{1}{4}\pi,\pi)$. The lower sign is chosen
\item When $c<0$ and $|c|<2a/\sigma^2$, $\text{ph}(\alpha_{1,2})|\in (\frac{1}{4}\pi,\frac{3}{4}\pi)$, the upper sign is chosen
\end{enumerate}

The choice is made based on the following properties of the complementary error function, as listed in ~\cite[Eq.~7.12.1]{NIST:DLMF}
\begin{enumerate}[label=(\alph*)]
\item $\erfc(z)e^{z^2}$ diverges in region $|\text{ph}(z)|\geq \frac{3}{4}\pi$.
\item $
\erfc(z)e^{z^2}\approx\frac{1}{\sqrt{\pi}}\sum_{m=0}^\infty (-1)^m \frac{\left(\frac{1}{2}\right)_m}{z^{2m+1}}$ for $ |\text{ph}(z)|<\frac{3}{4}\pi$
\item 
When $|\text{ph}(z)|<\frac{\pi}{2}$, the remainder terms are bounded by the first dropped terms times $\csc(2\text{ph}(z))$.
\item $\erfc(z)=2-\erfc(-z)$
\end{enumerate}
The first property is enough to fix the sign of the first situations, as the opposite sign makes the function divergent. Property (a) and (d) determines the sign of the second case. To see this, the real part of $\alpha_{1,2}/(1-i)$ is negative when $|y_2-y_1|<|c|\sigma^2/2-a$. This means $|\text{ph}(\alpha_{1,2})|\in (3\pi/4, \pi)$. To avoid divergence, the lower sign is chosen. At the same time $\erfc(z)\neq \erfc(-z)$ when $\Re(z)=0$. To make the wavefunction smooth, one need to impose the lower sign for all region of the arguments.  For the last scenario,the function behaves well with either sign, and one needs to take into account the rest of the properties. As we have shown, the attractive systems has bound states. These bound states should separate apart with the free state for large time. With our choice, when $t$ is large, $|\text{ph}(\alpha_{1,2})|\to 3\pi/4^-$. Thus, $\erfc(\alpha)e^{\alpha^2}\approx 2e^{\alpha^2}-1/(\sqrt{\pi}\alpha)$, where the first term corresponds to a bound state with factor $\exp(-|c|(|y_2-y_1|+a))$ and the second one is related to a free state which is identical to the repulsive solution. Note, when the attraction is too strong which corresponds to the second category, $|\text{ph}(\alpha_{1,2})|\to \pi/4$ when $t\gg 1$, thus there is no bound state in the systems.

Due to the complicated structure of the solution, it is difficult to obtain the density analytically except for a few limiting situations.

When $c=0$, we get
\begin{align}
f_{\uparrow,\downarrow}(y_1,y_2)=\frac{\sigma}{2\sqrt{\pi}i (t+\sigma^2/2i)}e^{\frac{i(y_2-x_1)^2}{4(t+\sigma^2/2i)}+\frac{i(y_1-x_2)^2}{4(t+\sigma^2/2i)}}
\end{align}
\begin{align}\label{eq:density1}
\langle\rho_{\uparrow}(y)\rangle=\frac{\sigma}{2\sqrt{\pi(t^2+\sigma^4/4)}}e^{-\frac{\sigma^2(y-x_1)^2}{4(t^2+\sigma^4/4)}}
\end{align}
\begin{align}\label{eq:density2}
\langle\rho_{\downarrow}(y)\rangle=\frac{\sigma}{2\sqrt{\pi(t^2+\sigma^4/4)}}e^{-\frac{\sigma^2(y-x_2)^2}{4(t^2+\sigma^4/4)}}
\end{align}
%\begin{align}\begin{split}
%\langle \rho_\uparrow(y_1)\rho_\downarrow(y_2)\rangle=\frac{\sigma^2}{4\pi(t^2+\sigma^4/4)}e^{-\frac{\sigma^2(y_2-x_1)^2}{4(t^2+\sigma^4/4)}-\frac{\sigma^2(y_1-x_2)^2}{4(t^2+\sigma^4/4)}}
%\end{split}\end{align}

\begin{figure}[h!]
\centering
	\begin{minipage}{0.5\textwidth}
		\centering
		\subfloat[][]
			{
				\includegraphics[width=.7\linewidth]{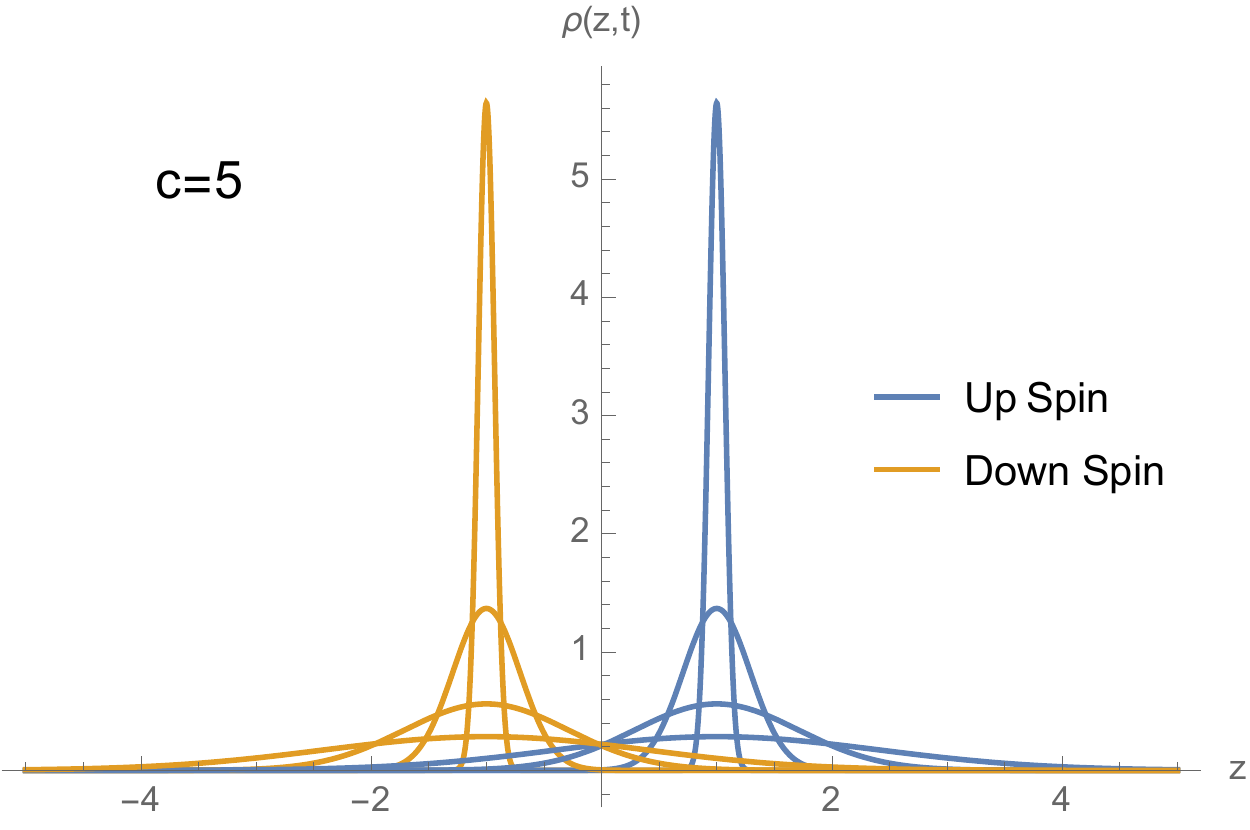}
				\label{GY_D1}
			}
			\\
	\subfloat[][]
		{
		 	\includegraphics[width=.7\linewidth]{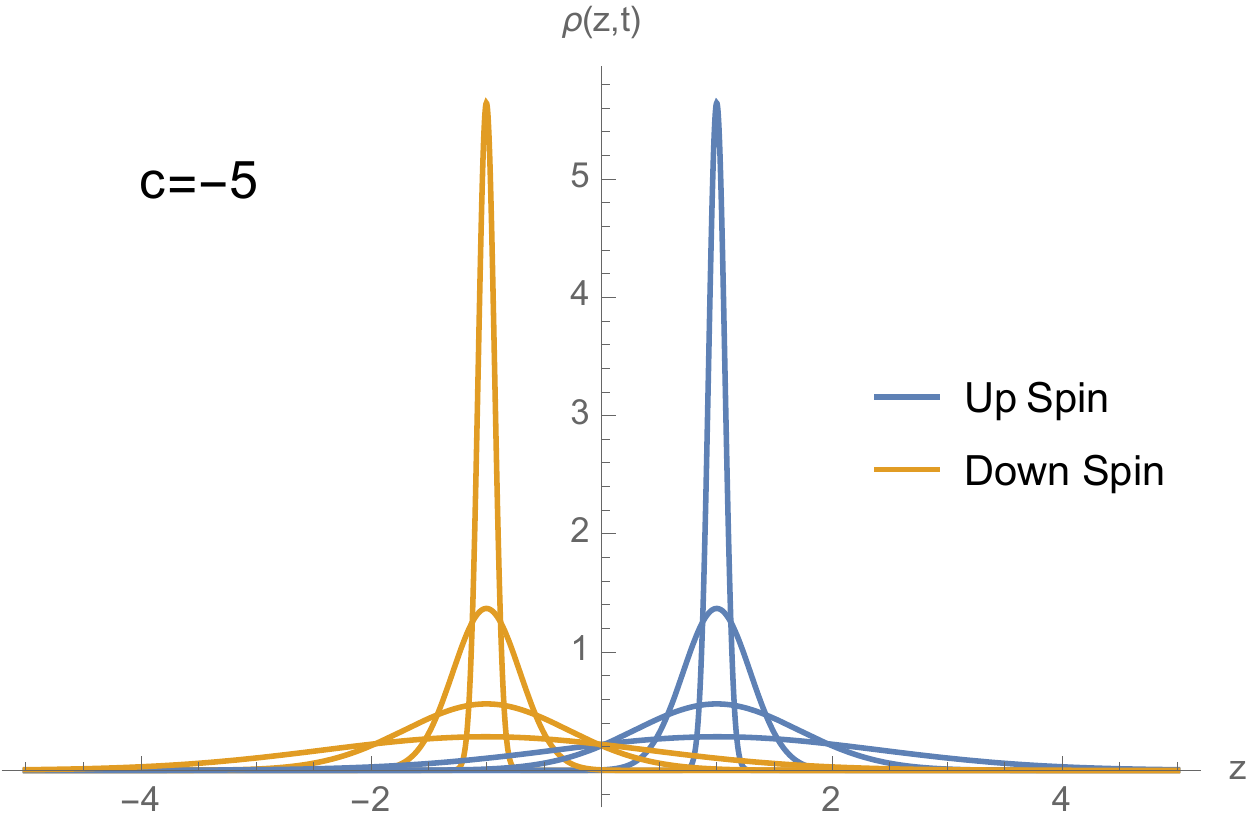}
			\label{GY_D2}
		}
	\\
	\subfloat[][]
		{
			\includegraphics[width=.7\linewidth]{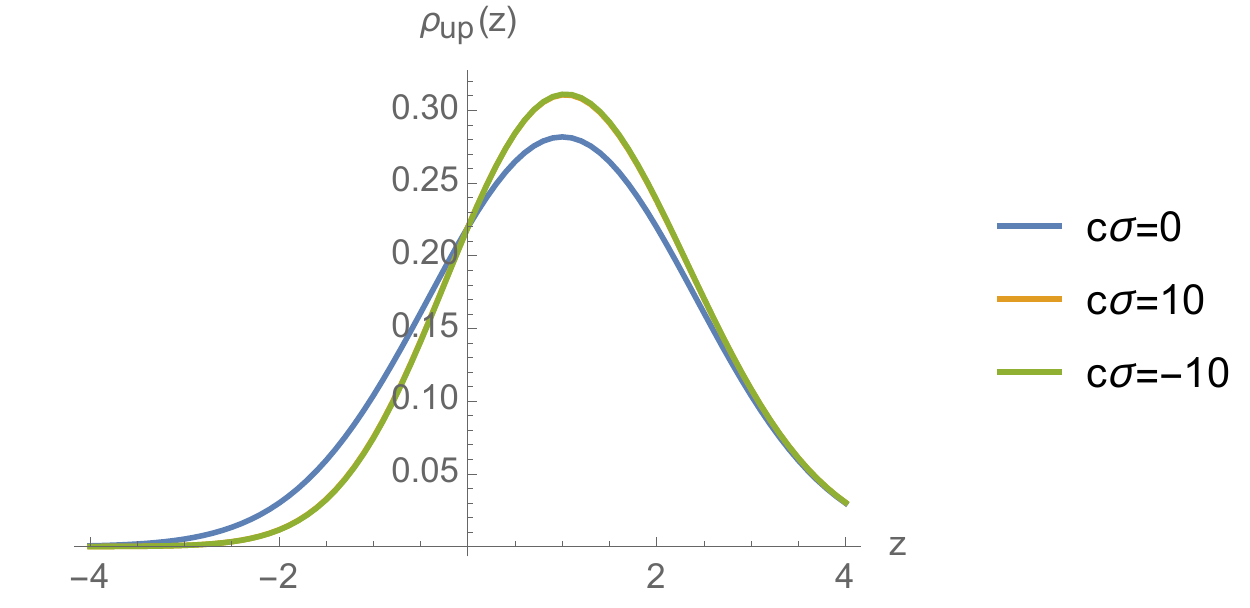}
			\label{GY_D4}
		}
\caption{Plot a and b show time evolution of down spin (yellow) and up spin (blue) at $t=0, 0.02, 0.05, 0.1$ for $c=5$ and $c=-5$ respectively. Figure c compares the density of the up spin for different interactions.}
\label{fig:density}
\end{minipage}
\end{figure}

When $c=\pm \infty$, we get\footnote{Note, Mathematica fails for produce the correct answer for $\rho_{\downarrow}(y)$ by performing the integration analytically}.
\begin{align}\begin{split}
f_{\uparrow,\downarrow}(y_1,y_2)=&\frac{\sigma}{2\sqrt{\pi}i (t+\sigma^2/2i)}\big(e^{\frac{i(y_2-x_1)^2}{4(t+\sigma^2/2i)}+\frac{i(y_1-x_2)^2}{4(t+\sigma^2/2i)}}\\
&-e^{\frac{i(y_1-x_1)^2}{4(t+\sigma^2/2i)}+\frac{i(y_2-x_2)^2}{4(t+\sigma^2/2i)}}\big) \theta(y_1-y_2)
\end{split}\end{align}
\begin{align}\begin{split}
&\rho_{\uparrow}(y)\\
&=\frac{\sigma}{4\sqrt{\pi (t^2+\sigma^4/4)}}\big(e^{-\frac{\sigma^2(y-x_2)^2}{4(t^2+\sigma^4/4)}}\erfc(\frac{\sigma (x_1-y)}{2\sqrt{t^2+\sigma^4/4}})\\
&-e^{\frac{i(y-x2)^2}{4(t+\sigma^2/2i)}-\frac{i(y-x1)^2}{4(t-\sigma^2/2i)}}\erfc(\frac{\frac{x_1+x_2-2y}{2}\sigma-\frac{it(x_1-x_2)}{\sigma}}{2\sqrt{t^2+\sigma^4/4}})\big)\\
&\times e^{-\frac{(x_1-x_2)^2}{4\sigma^2}}+(x_1\leftrightarrow x_2)
\end{split}\end{align}

\begin{align}\begin{split}
&\rho_{\downarrow}(y)\\
&=\frac{\sigma}{4\sqrt{\pi (t^2+\sigma^4/4)}}\big(e^{-\frac{\sigma^2(y-x_2)^2}{4(t^2+\sigma^4/4)}}\erfc(\frac{\sigma (y-x_1)}{2\sqrt{t^2+\sigma^4/4}})\\
&-e^{\frac{i(y-x1)^2}{4(t+\sigma^2/2i)}-\frac{i(y-x2)^2}{4(t-\sigma^2/2i)}}\erfc(\frac{\frac{2y-x_1-x_2}{2}\sigma-\frac{it(x_1-x_2)}{\sigma}}{2\sqrt{t^2+\sigma^4/4}})\big)\\
&\times e^{-\frac{(x_1-x_2)^2}{4\sigma^2}}+(x_1\leftrightarrow x_2)
\end{split}\end{align}
For general interactions, the density are plotted in Figure~\ref{fig:density}. As shown in~\ref{GY_D1} and~\ref{GY_D2}, both repulsive and attractive systems shows Guassian diffusion similar to the free model described by equation~(\ref{eq:density1},\ref{eq:density2}). In Figure~\ref{GY_D4}, we compared the density distribution of the up spin for different interaction strengths. We saw that the shape of the density depends only on the strength of the interaction, not the sign of it. The reason for a similar behavior between systems with $c>0$ and $c<0$ is related to energy conservation and the initial state we have chosen. Since the particles have negligible overlaps among each other right after the quench, the system have little interaction energy at the beginning. When the particles are allowed to expand freely, they still avoid contact among each other as it will leads to decrease or increase of the total energy. 

\begin{figure}[h!]
\centering
	\begin{minipage}{0.5\textwidth}
		\centering
		\subfloat[][]
			{		
				\includegraphics[width=0.7\linewidth]{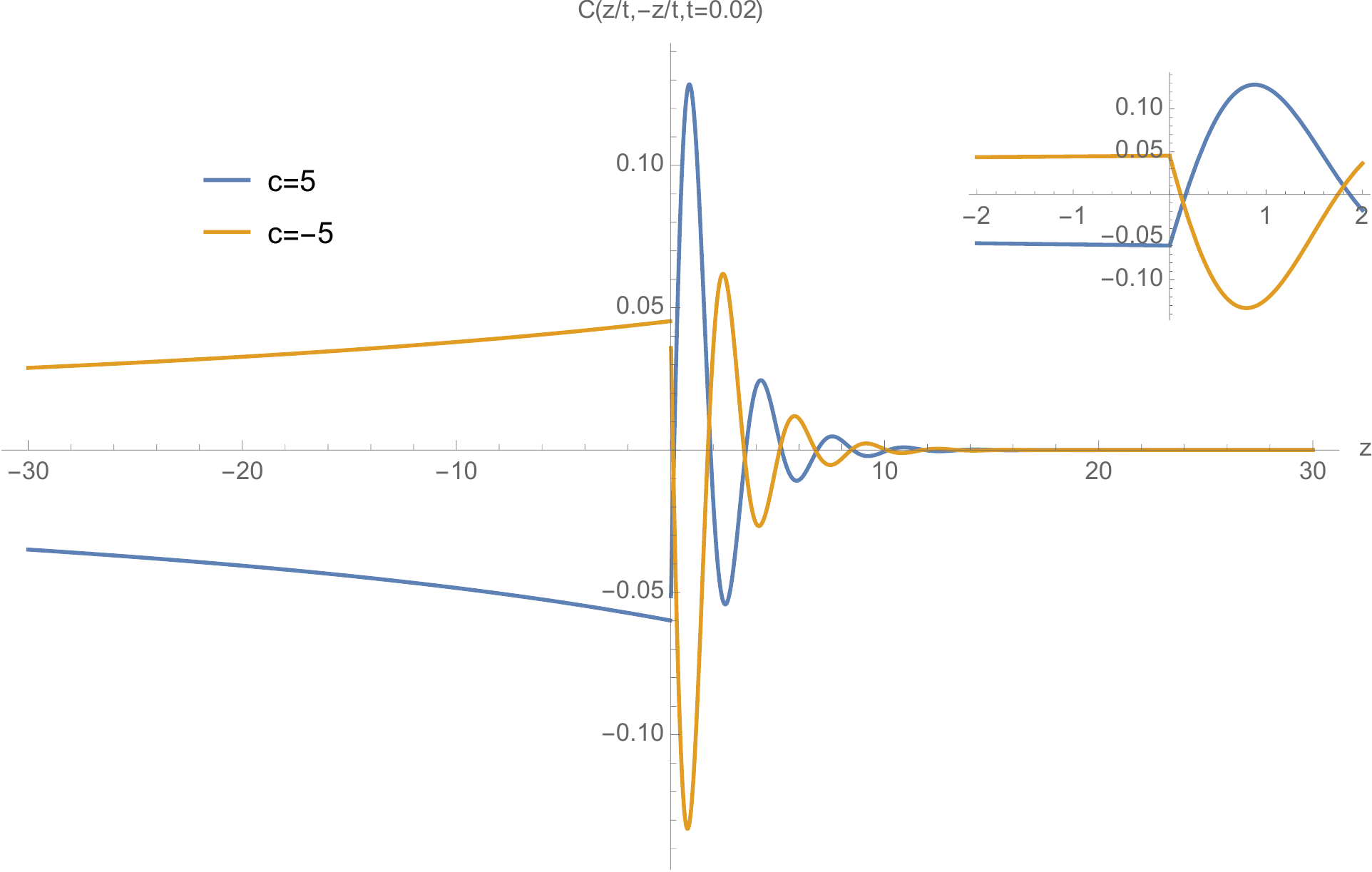}
				\label{GY_Cr_1}
			}
		\\
		\subfloat[][]
		{
			\includegraphics[width=0.7\linewidth]{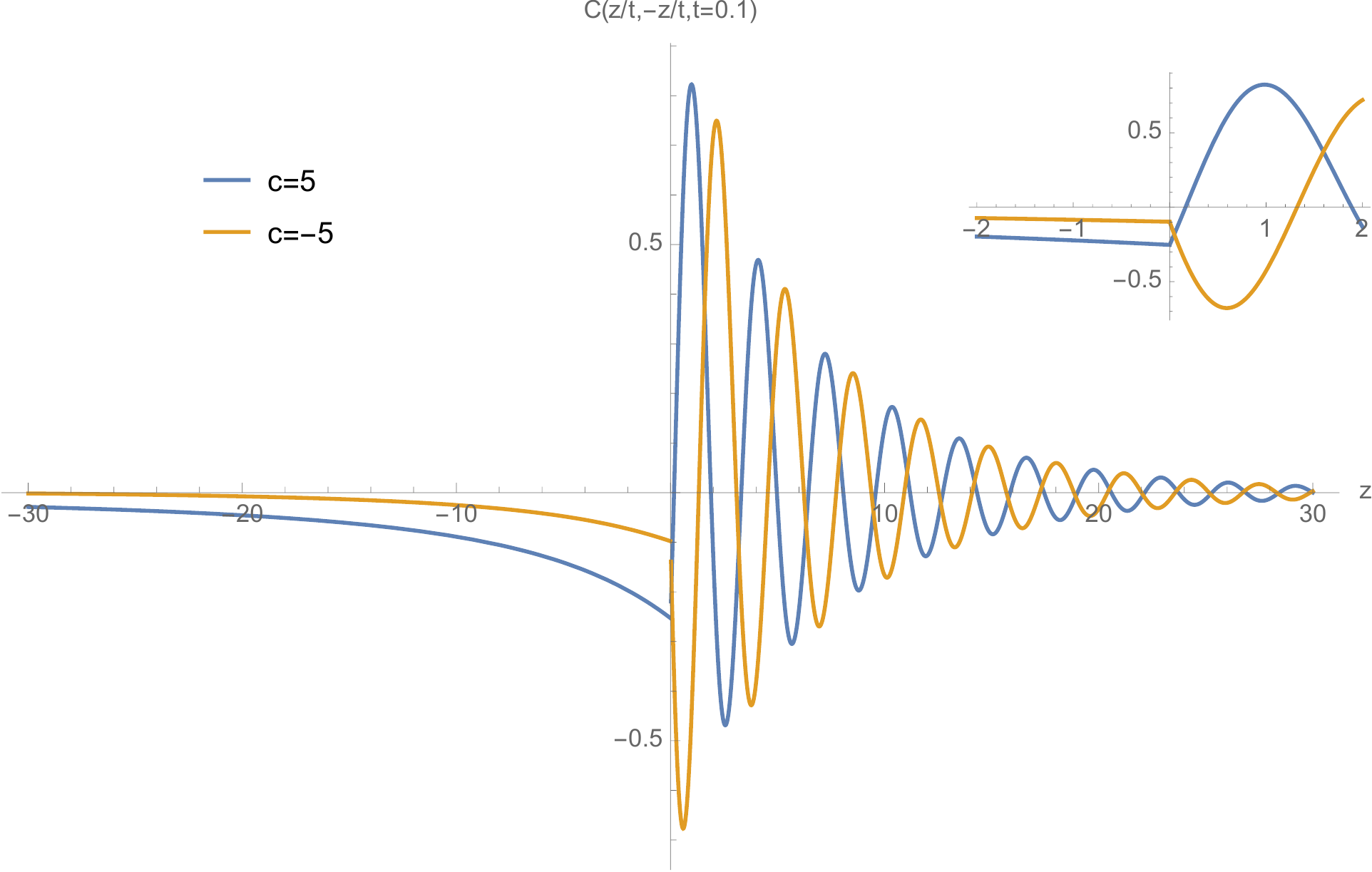}
			\label{GY_Cr_2}
		}
		\\
		\subfloat[][]
		{
			\includegraphics[width=0.7\linewidth]{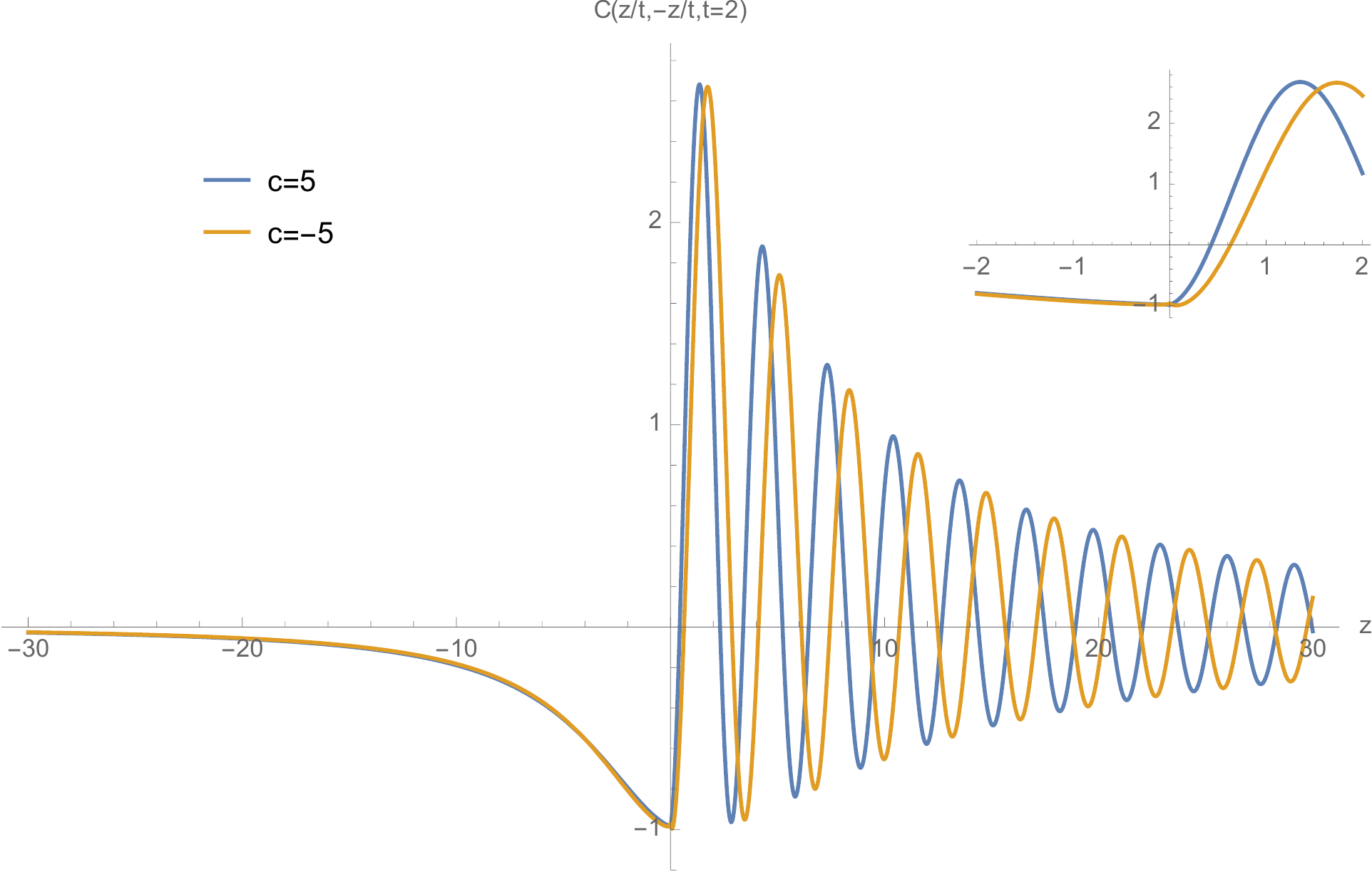}
			\label{GY_Cr_4}
		}
\caption{Normalized noise correlation for a system with one up spin and one down spin with static impurity initial state. The down spin is initially to the left to the up spin. The figure shows the correlation at $t=0.02$~\protect\subref{GY_Cr_1}, $0.1$~\protect\subref{GY_Cr_2} and $2$~\protect\subref{GY_Cr_4} respectively. The insets show details near the origin.}
\label{fig:noise}
\end{minipage}
\end{figure}

\begin{figure}
	\centering
	\includegraphics[width=0.45\textwidth]{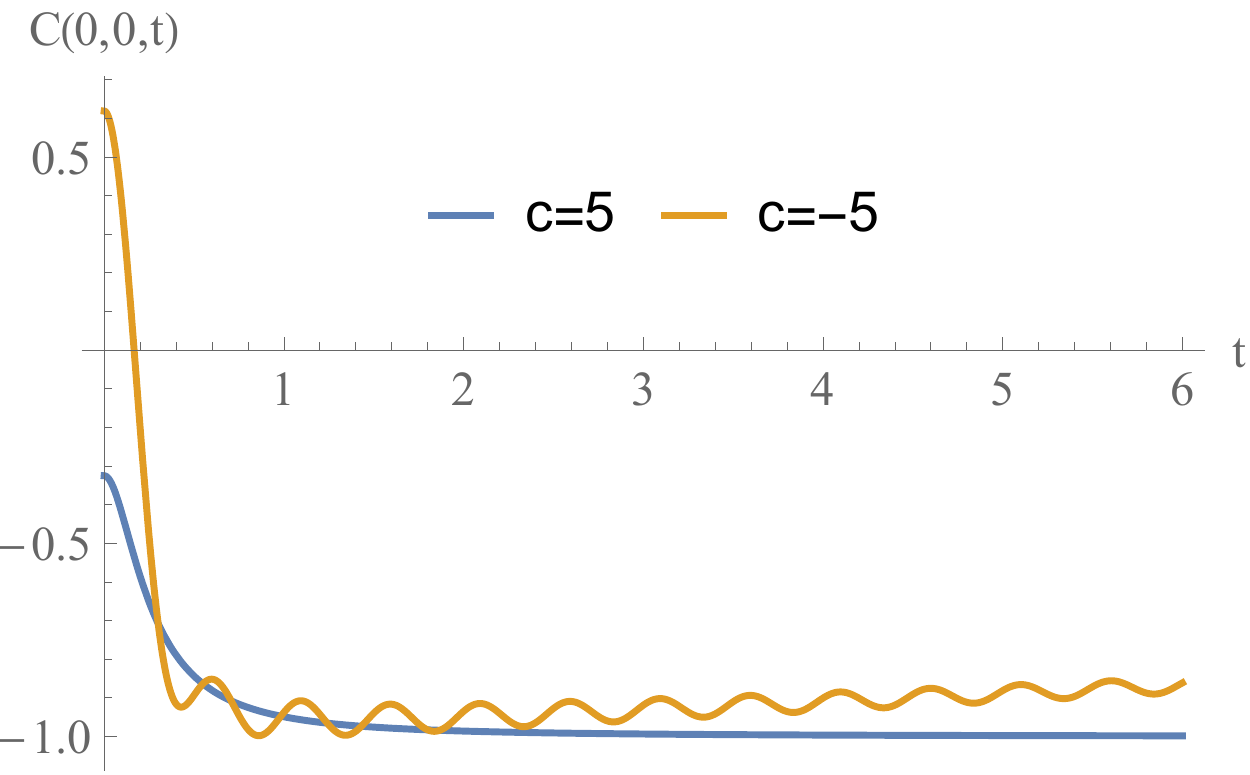}
	\caption{Unnormalized correlation function at the origin as function of time. Blue line is for repulsive and yellow line applies to attractive case.}
	\label{corr0}
\end{figure}

This picture is further confirmed by our calculation of the normalized noise function, $C(z/t,$ $-z/t,t)=\frac{\langle \rho_{\uparrow}(z/t)\rho_{\downarrow}(-z/t)\rangle}{\langle \rho_{\uparrow}(z/t)\rangle \langle\rho_{\downarrow}(-z/t)\rangle}-1$=$\langle\frac{\delta\rho_{\uparrow}(z/t)\delta\rho_{\downarrow}(-z/t)\rangle}{\langle \rho_{\uparrow}(z/t)\rangle \langle\rho_{\downarrow}(-z/t)\rangle}$, which are shown in Figure~\ref{fig:noise}. When $t=0.02$, the correlation at the origin is positive with attractive interaction and is negative with repulsive interaction. However, the correlations for both cases approach $-1$, i.e. $\rho_\uparrow(0)\rho_\downarrow(0)\to 0$, as time evolves. Figure \ref{corr0} compares the correlation function at different times. Indeed, we saw that shortly after the quench, difference in the correlations is substantial. Then both correlations approach $-1$ quickly. Moreover,  when time gets greater, the attractive correlation function gradually increases. This is due to the fact that bound states diffuse slower, thus their contribution is more prominent in the asymptotic limit where overlap is little.


% Away from the origin, we see oscillations on the positive side, which results from the superposition of two terms, one with no crossing or spin exchange, one with both of them. On the negative side, there is only one term that leads to the down spin lying to the right of the up spins. So the noise function does not oscillate. 

\subsection{Dynamics of One Impurity in $N-1$ Fermion Bath}

\begin{figure}[hb]
\centering
	\begin{minipage}{0.45\textwidth}
	\centering
	\subfloat[][]
	{
		\includegraphics[width=1\linewidth]{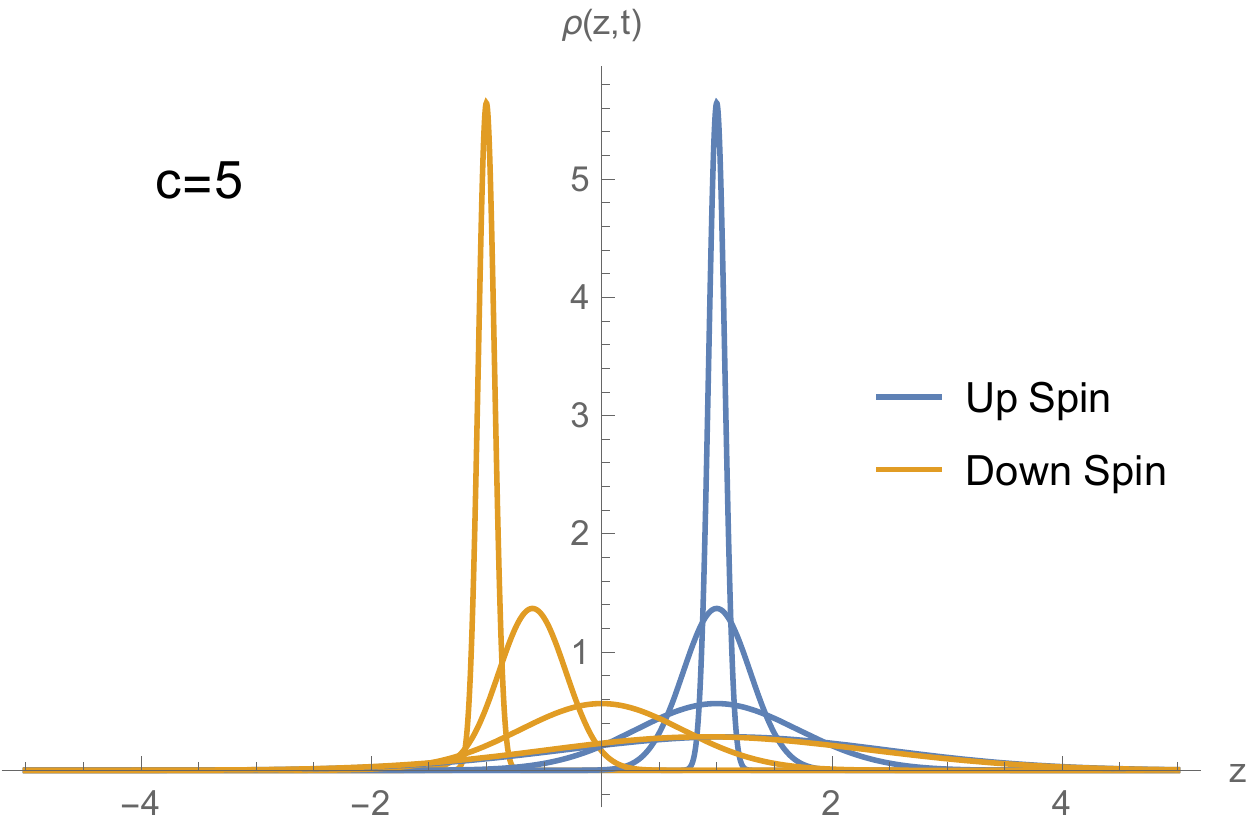}
		\label{Moving_1a}
	}\\
	\subfloat[][]
	{
		\includegraphics[width=1\linewidth]{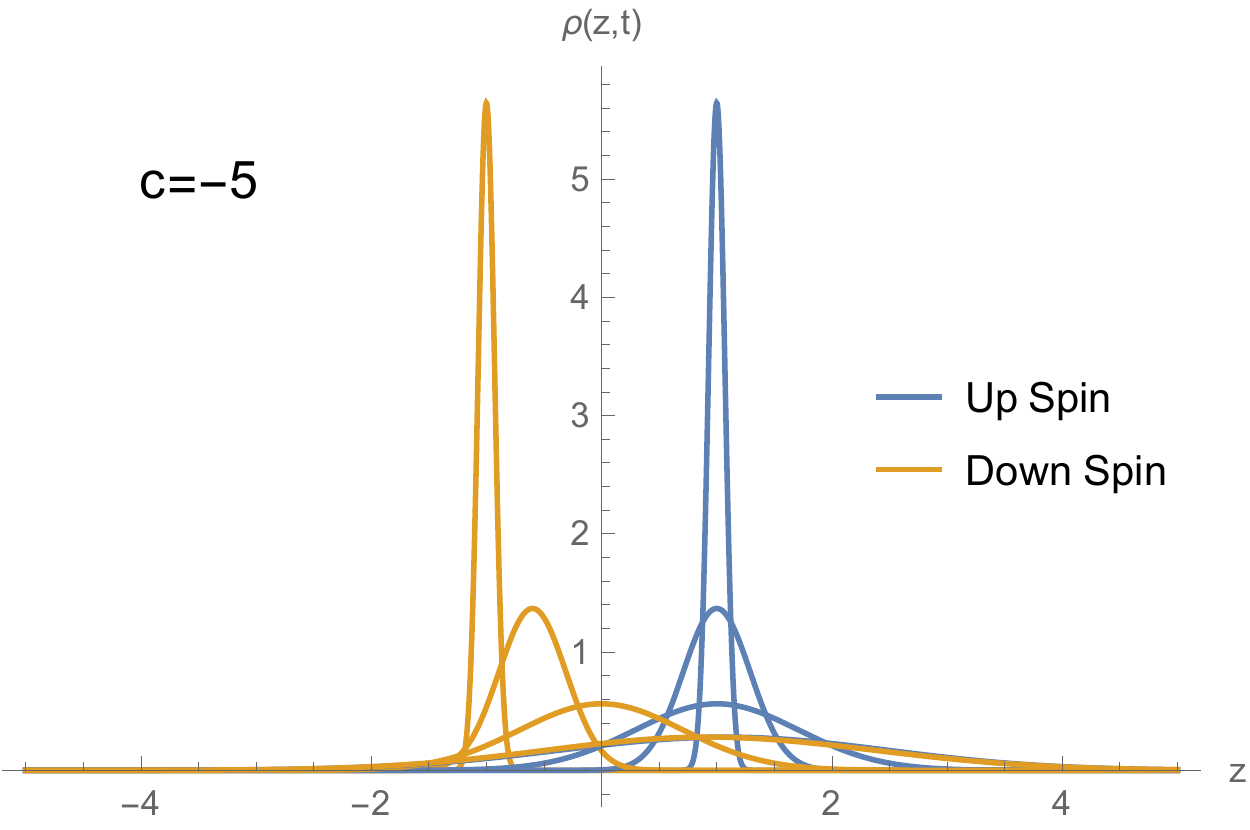}
		\label{Moving_1b}
	}
	\caption{Figure shows the time evolution of the down spin (yellow) and the up spin (blue) at $t=0,0.02,0.05,0.1$ for systems with $N=2$ and $M=1$. At $t=0$, the down spin moves towards the up spin with momentum $k_0=1$. \protect\subref{Moving_1a} is for repulsive case and \protect\subref{Moving_1b} is for attractive case.}
	\label{Moving_1}
	\end{minipage}
\end{figure}

In this section, we discuss the time evolution of systems with $N-1$ fermions with up spin and $1$ fermion with down spin (the impurity). We study the time evolution of an initial state with either static or kinetic impurity. The initial state can be written as
\begin{align}\begin{split}
|\phi_0(x,\beta,k_0)\rangle=\frac{1}{(\pi\sigma^2)^{N/4}}\int_{x'} e^{-\sum_i^N\frac{(x'_i-x_{i})^2}{2\sigma^2}+ik_0 x'_\beta}|x',\beta\rangle
\end{split}\end{align}
with $x_{i+1}-x_i>3\sigma$. Here the impurity fermion has an initial momentum $k_0$. When $k_0=0$, we get the static impurity problem. Note, the new phase factor $e^{ik_ox'_\beta}$ does not affect the proof of the central theorem. As the latter only involves $k$ and $\mu$ integration, where the new phase is simply a constant. The two particle wavefunction Eq.~(\ref{2w}) is modified as $f_{\uparrow,\downarrow}(y_1,y_2, k_0)=f_{\uparrow,\downarrow}(y_1,y_2)_{x_1\to x_1+ik_0\sigma^2}\exp(ik_0\sigma^2-k_0^2\sigma^2/2)$. The result are shown in Figure \ref{Moving_1}. Again, no significant difference appears between systems with attractive and repulsive interaction.

Unlike the two-particle case, when $N>2$, the $k$ integration cannot be carried out explicitly. We make asymptotic simplification as in~\cite{deepak1}, and replace the integral by their saddle point contribution at $k_i=\xi_{\inv{P}i}-ik_0\sigma^2/2t\delta_{i\beta}$ with $\xi_i=y_i/2t$. Here we have dropped terms in higher order of $x_i/t$ and $\sigma^2/t$.  Depending on the sign of the interaction, the asymptotic wavefunction gets different forms. As we will discuss separately.
\subsubsection{Repulsive Interaction}

When the interaction is repulsive, one can shift all contours to the real axis without crossing any pole. Then the wavefunction becomes
\color{black}
\begin{align}\begin{split}
&f(\xi,\alpha,t)\\
=&\frac{\sigma^{\frac{N}{2}}}{2^{\frac{N}{2}+1}\pi^{\frac{N}{4}+1}c(it)^{\frac{N}{2}}} e^{\sum_i it\xi_i^2-\xi_i^2\sigma^2/2-i\xi_i x_{Pi}-\sigma^2\xi_{\inv{P}\beta}k_0}(-1)^P \\
&\int_\mu J(\mu)\theta(\xi_1<\ldots\xi_N)\\
=&\frac{i\sigma^{\frac{N}{2}}}{2^{\frac{N}{2}}\pi^{\frac{N}{4}}c(it)^{\frac{N}{2}}}e^{\sum_i it\xi_i^2-\xi_i^2\sigma^2/2-i\xi_i x_{Pi}-\sigma^2\xi_{\inv{P}\beta}k_0}(-1)^P \sum_o\\
& R(\xi_o+ic/2) \theta(\xi_1<\ldots\xi_N)
\end{split}\end{align}
$R(\xi+ic/2)$ being the residue of $J(\mu)$ at $\xi+ic/2$ as introduced in section~\ref{sec: single}. The expression summed over all poles that lies above $\mu$ contour. Density and noise function can be calculated from it as
\begin{align}
\rho_{\uparrow}(z)=(2t)^{N-1}\int_\xi\sum_{\alpha} \sum_{i\neq \alpha}|f(\xi,\alpha,t)|^2\delta(\xi_i-z/2t)
\end{align}
\begin{align}
\rho_{\downarrow}(z)=(2t)^{N-1}\int_\xi \sum_{\alpha}|f(\xi,\alpha,t)|^2\delta(\xi_\alpha-z/2t)
\end{align}
\begin{align}
\begin{split}
&\rho_{\downarrow\uparrow}(z,z')\\
&=(2t)^{N-2}\int_\xi \sum_{\alpha}\sum_{i\neq \alpha}|f(\xi,\alpha,t)|^2\delta(\xi_\alpha-z)\delta(\xi_i-z')
\end{split}
\end{align}
\begin{align}
C(z,z',t)=\frac{1}{N-1}\frac{\rho_{\downarrow\uparrow}(z,z')}{\rho_\downarrow(z) \rho_\uparrow(z')}-1
\end{align}
\begin{align}
\begin{split}
&|f(\xi,\alpha,t)|^2=\frac{\sigma^N }{2^N\pi^{\frac{N}{2}} t^{N}c^2}\theta(\xi_1<\ldots<\xi_N)e^{-\sum_i\sigma^2\xi_i^2}\\
&\ \ \ \times \sum_{P,P'}(-1)^{P+P'}e^{-i\sum_i\xi_i (x_{Pi}-x_{P'i})-k_0\sigma^2(\xi_{\inv{P}\beta}+\xi_{\inv{P'}\beta})}\\
&\ \ \ \times \sum_{o,e}R(\xi_o+ic/2)R^*(\xi_e-ic/2)
\end{split}
\end{align}
In order to obtain observables, one needs to integrate out dummy variables in the above expression. First, we impose the condition that $x_{Pi}=x_{P'i}$ if $\xi_i$ is integrated over without $\delta$ function. The oscillation related to these terms makes the contribution small by a factor of $e^{-|c|a}$. Moreover, the number of terms from $R(\xi_o+ic/2)R^*(\xi_e-ic/2)$ is as many as $N^2 (N!)^2$. It becomes difficult to keep track of all of them when the number of particles is large. To make the calculation tractable, we only keep the terms in which $R(\xi_o+ic/2)R^*(\xi_o-ic/2)$ does not depend on any dummy variables. This leads to approximations in the leading order of $c\sigma$. We have checked the contribution of these small terms in a small system ($N=3$). These dropped terms turn out to be small up to $(c\sigma)^4$. In the following calculations, we will exploit these two simplifications.

\color{black}For density function, the leading order term comes from $e=o=\alpha=\invbig{P}\beta$ when $R(\xi_o+ic/2)=-ic$. All other terms depends on at least two variables.  Moreover, $Pi=P'i$ for any $i\neq \alpha$ indicates that $P=P'$. Thus we have
\begin{align}
\begin{split}
\rho_\downarrow(z)=&\frac{\sigma^N}{2^N \pi^{N/2}t^N c^2} (2\pi c)^2\sum_P\int_{\xi} e^{-\sigma^2\sum_i\xi_i^2-2k_0\sigma^2\xi_\alpha}\\
&e^{-k_0^2\sigma^2}\theta(\xi_1<\ldots<\xi_N)\delta(\xi_\alpha-z/2t)\\
=&\frac{\sigma}{2\sqrt{\pi}t}\exp(-\sigma^2(z-2tk_0)^2/4t^2)
\end{split}
\end{align}

Similarly, we have $\rho_{\uparrow}(z)=(N-1)\sigma/(2\sqrt{\pi}t)$ $\exp(-\sigma^2z^2/4t^2)$. Thus, the leading order behavior of the density is Gaussian diffusion as in free models.

To calculate correlation functions,  we assume that $x_{Pj}=x_{P'j}$ for $j\neq \alpha$ or $i$. Moreover, only terms in $R(\xi_o+ic/2)R^*(\xi_e-ic/2)$ that do not depend these $\xi_j$ are included. These terms are
\begin{align}\label{drop}
\begin{split}
&R(\xi_o+ic/2)\rightarrow\\
&-ic\delta_{P\alpha,\beta}\big(1+\theta(\alpha-i)\theta(Pi-\beta)\frac{-ic}{\xi_\alpha-\xi_i+ic}+\theta(\beta\\
&\ -Pi)\theta(i-\alpha)\frac{-ic}{\xi_i-\xi_\alpha+ic}\big)+ic\delta_{\beta,Pi}\big(\theta(\invbig{P}\beta-\alpha)\\
&\times \theta(P\alpha-\beta)\frac{-ic}{\xi_{\inv{P}\beta}-\xi_\alpha+ic}+\theta(\alpha-\invbig{P}\beta)\theta(\beta-P\alpha)\\
&\times \frac{-ic}{\xi_\alpha-\xi_{\inv{P}\beta}+ic}\big)
\end{split}
\end{align}
This leads to the following noise function
\begin{align}\label{RNoise}
\begin{split}
&C(z,z',t)\\
=&\frac{\theta(z-z')}{N-1}\Big((1+\frac{4t^2c^2}{(z-z')^2+4t^2c^2}e^{\frac{\sigma^2k_0(z'-z)}{t}})(\beta-1)\\
&+(N-\beta)\frac{(z-z')^2}{(z-z')^2+4t^2c^2}-2\Im(\frac{2tc}{z-z'-2ict}\\
&\times e^{-\frac{\sigma^2k_0(z-z')}{2t}}\frac{e^{-\frac{i(z-z')a}{2t}}-e^{-i\frac{(z-z')\beta a}{2t}}}{1-e^{-\frac{i(z-z')a}{2t}}})\Big)+\frac{\theta(z-z')}{N-1}\\
&\Big((1+\frac{4t^2c^2}{(z-z')^2+4t^2c^2}e^{\frac{\sigma^2k_0(z'-z)}{t}})(N-\beta)+(\beta-1)\\
&\times\frac{(z-z')^2}{(z-z')^2+4t^2c^2}-2e^{-\frac{\sigma^2k_0(z-z')}{2t}}\Im(\frac{2tc}{z-z'-2ict}\\
&\times\frac{e^{-\frac{i(z-z')a}{2t}}-e^{-i\frac{(z-z')(N-\beta+1) a}{2t}}}{1-e^{-\frac{i(z-z')a}{2t}}})\Big)-1\\
\end{split}
\end{align}

\begin{figure}[hb]
\centering
%	\begin{minipage}{0.45\textwidth}
%	\centering
			\subfloat[][]
			{
			\includegraphics[width=0.45\linewidth]{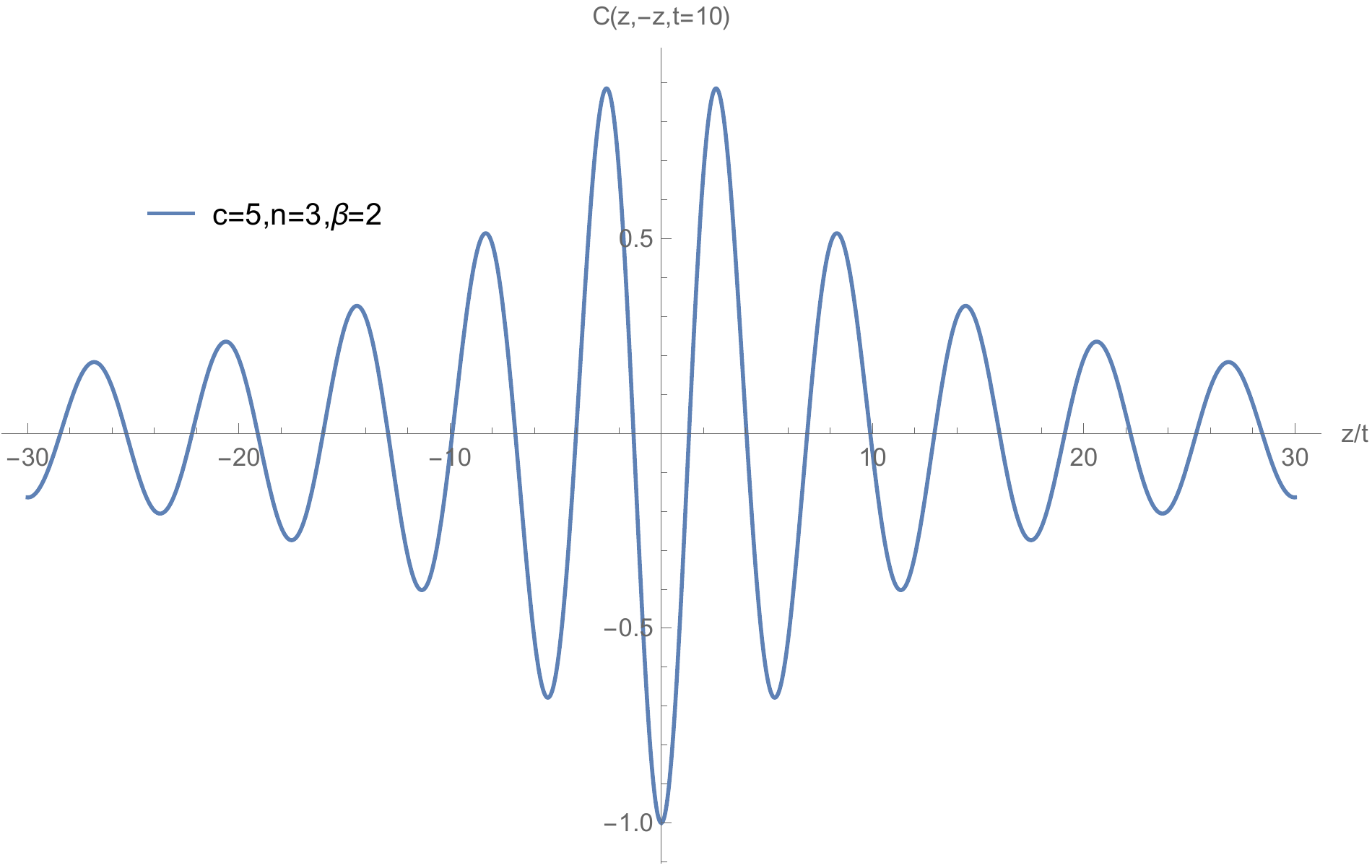}
			\label{GY_noise_n3}
			}
		\subfloat[][]
			{
				\includegraphics[width=0.45\linewidth]{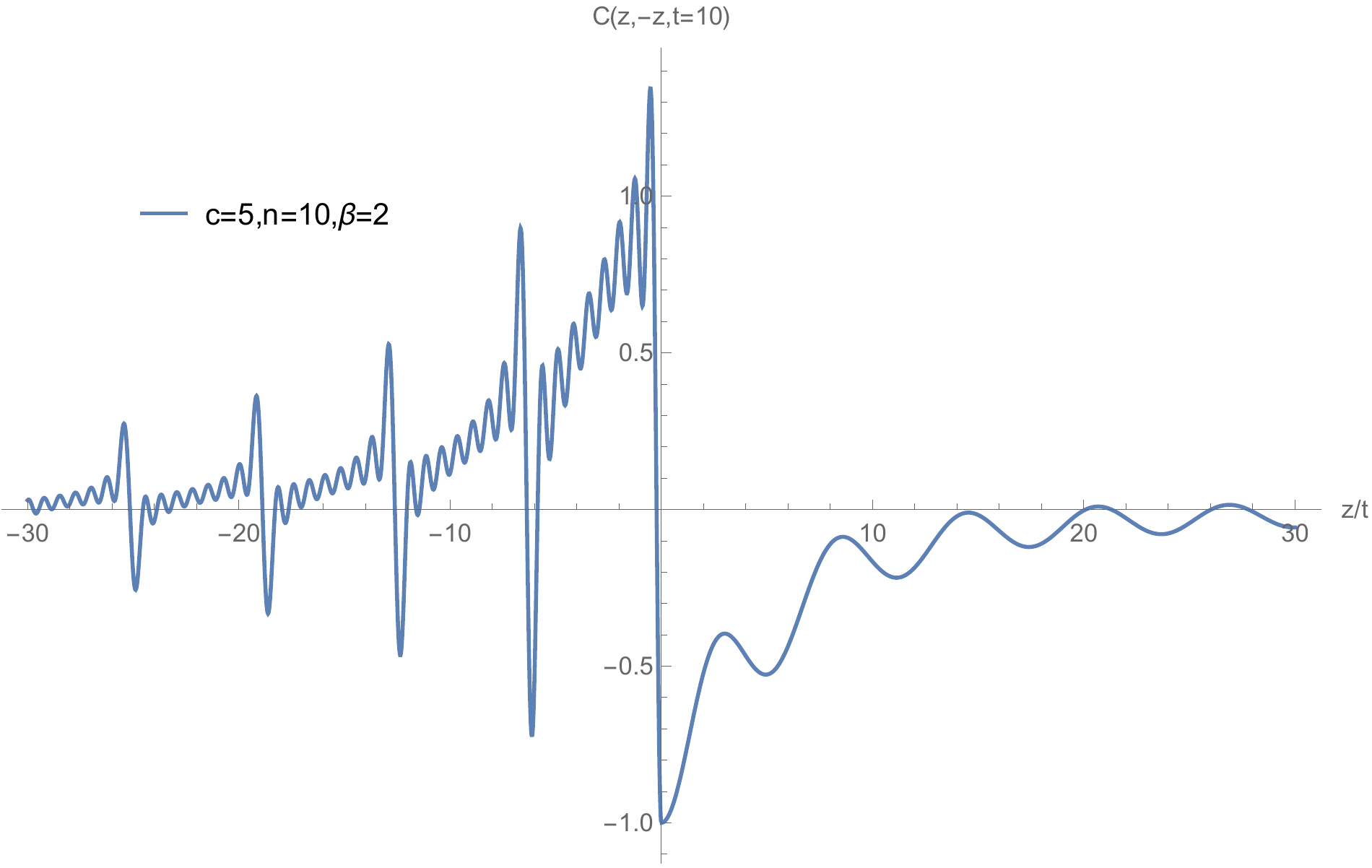}
				\label{GY_noise_n10}
			}
		\\
		\subfloat[][]
			{
			\includegraphics[width=0.45\linewidth]{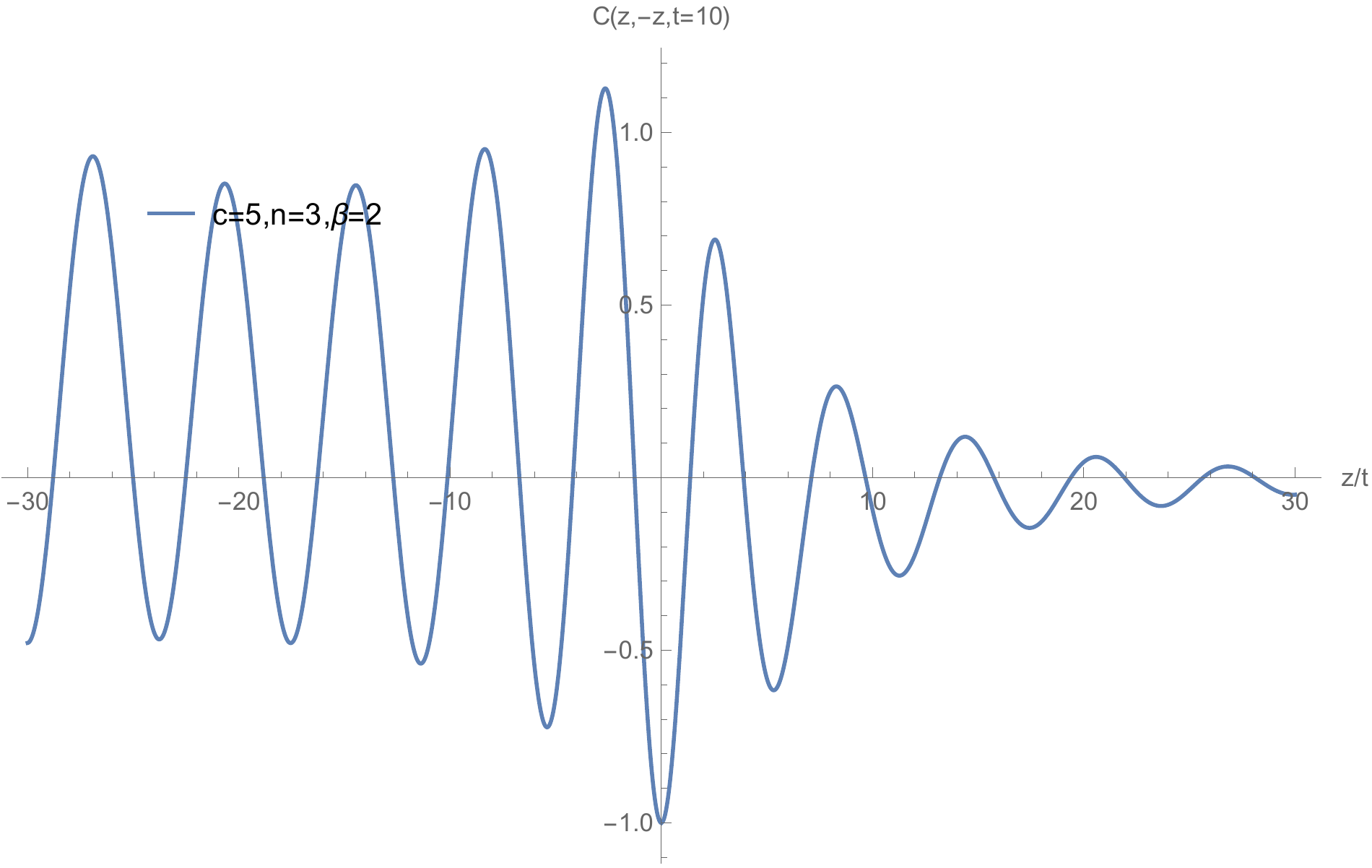}
			\label{GY_noise_k_n3}
			}
		\subfloat[][]
			{
				\includegraphics[width=0.45\linewidth]{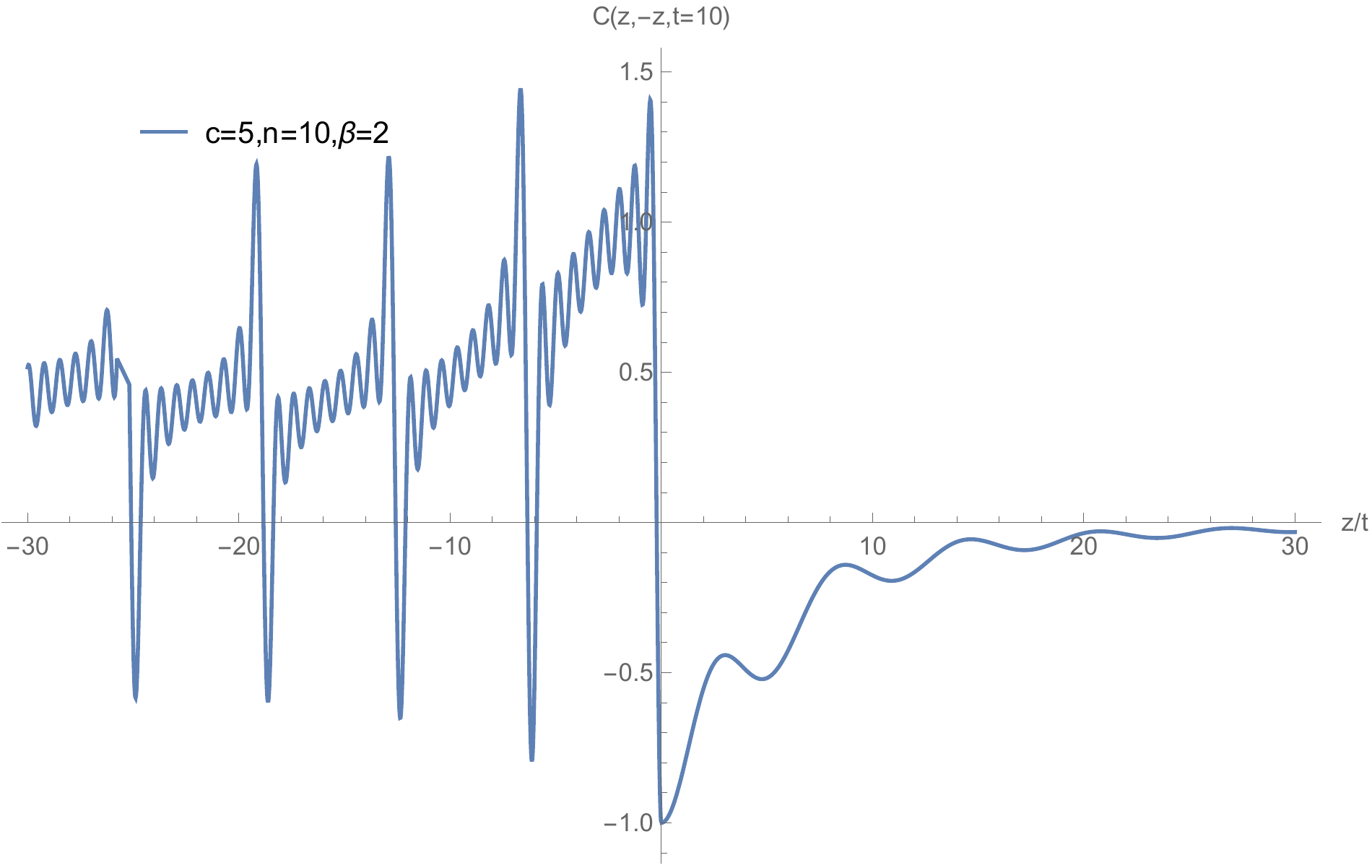}
				\label{GY_noise_k_n10}
			}
\caption{The plots show the noise function for static~(Fig.\ref{GY_noise_n3},\ref{GY_noise_n10}) and kinetic~(Fig. \ref{GY_noise_k_n3},\ref{GY_noise_k_n10}) impurity. Figure  \ref{GY_noise_n3} and Figure \ref{GY_noise_k_n3} are for systems with $N=3$ fermions with the impurity sitting in the middle. Figure \ref{GY_noise_n10} and Figure \ref{GY_noise_k_n10} are related to systems with $N=10$ fermions, and $\beta=2$.}
\label{GY_noise_k}
%\end{minipage}
\end{figure}
Figure \ref{GY_noise_k} shows the result for static (Fig. \ref{GY_noise_n3}, \ref{GY_noise_n10}) and kinetic~(Fig. \ref{GY_noise_k_n3}, \ref{GY_noise_k_n10}) impurity cases. Figure  \ref{GY_noise_n3} and Figure \ref{GY_noise_k_n3}) are for systems with $N=3$ $\beta=2$. The noise function shows oscillations with period $t/a$. Figure \ref{GY_noise_n10} and \ref{GY_noise_k_n10} show noise function for $N=10$ fermions with $\beta=2$. The envelop of the noise function shows periodic dips at the same position as in the previous case. As discussed in~\cite{rom2006free}, their position is purely a statistical effect. Superimposed on these patterns are finer interference fringes on the negative side. This is due to all possibilities that leads to an up spin lying to the left of a down spin. The number of such fringes equals $N-\beta$, or $\beta-1$ for fringes on the right. At the origin, the noise function equals $-1$, indicating that it is impossible to have the up spin and down spin to occupy the same position. This is due to the same reason as the two particle case, i.e. to avoid energy change caused by contact interaction. This is true even for the kinetic impurity example. That is to say, the system does not pay the change of interaction energy by tuning the kinetic energy of the impurity. Since the noise function for any eigenstate, free or bound state, is greater than $-1$, such behaviour near the origin indicates a superposition of many eigenstates.

\subsubsection{Attractive Interaction}
As we have shown in section \ref{sec:bs}, unlike in the repulsive cases, bound states exist in attractive models. These bound states manifest themselves as pole contributions in the wavefunction. Thus, the wavefunction can be split into free state terms and bound state terms. 
\begin{align}
\begin{split}
&f(\xi,x,t)\\
=&\frac{\sigma^{N/2} i}{2^{N/2}\pi^{N/4}c(it)^{N/2}}\ e^{\sum_j i(t-\sigma^2/2i)\xi_{\inv{P}j}-i\xi_{\inv{P}j}(x_j+ik_o\sigma^2\delta_{j\beta}}\\
&e^{ik_0x_0-k_0^2\sigma^2/2}\sum_{\substack{o\leq\beta \inv{P}0\geq \alpha}}R(k_o+ic/2)\\
-&\frac{\sigma^{N/2}i}{2^{(N+1)/2}\pi^{N/4+1/2}(it)^{(N-1)/2}}\sum_{\substack{o\leq\beta\leq Pm\\
m\leq\alpha\leq \inv{P}o}}e^{\sum_{j} i(t-\sigma^2/2i)\xi_{\inv{P}j}}\\
&e^{-i\xi_{\inv{P}j}(x_j
+ik_o\sigma^2\delta_{j\beta})
+ik_0x_0-\frac{k_0^2\sigma^2}{2}+\frac{i(\xi_{\inv{P}o}+\xi_m)^2(t-\sigma^2/2i)}{2}-ct\xi_m}\\
&e^{ct\xi_{\inv{P}o}-\frac{i(\xi_{\inv{P}0}+\xi_m)(x_o+x_{Pm}+ik_0\sigma^2\delta_{o\beta}+ik_0\sigma^2\delta_{Pm,\beta})}{2}+\frac{ick_0\sigma^2}{2}(\delta_{Pm\beta}}\\
&e^{-\frac{ick_0\sigma^2}{2}\delta_{o\beta}+\frac{c(x_{Pm}-x_o)}{2}+\frac{ic^2(t+\sigma^2/2i)}{2}}R(R(\mu=k_0+ic/2)\\
&,k_o=k_{Pm}-ic))
\end{split}
\end{align}
In the above expression, all $k$'s should be replaced by their saddle point value. For the first term, that corresponds to 
\[
k_i=-(y_{\inv{P}i}-x_i)/2t
\]
For the second term, that corresponds to 

\begin{equation*}
k_i=\begin{cases}
	(y_{\inv{P}i}-x_i)/2t& i\neq Pm, o \\
  	(y_m-x_{Pm}+y_{\inv{P}o}-x_o)/4t-ic/2 & i=o\\
  	(y_m-x_{Pm}+y_{\inv{P}o}-x_o)/4t+ic/2 & i=Pm
  \end{cases}
\end{equation*}

Note, a single $k$ may depend on $y_m$ and $y_{\inv{P}o}$ simultaneously. Now use the same approximation as we did for the repulsive case, the leading order contribution in $R(R(\mu=k_0+ic/2),k_o=k_{Pm}+ic/2))$ for both density and noise function comes from
\begin{align}
\begin{split}
\text{R}\rightarrow& (ic)^2\delta_{P\alpha\beta}\delta_{P\alpha o}\theta(\alpha-m)\theta(Pm-P\alpha)
\\-&(ic)^2 \delta_{m\alpha}\delta_{\beta  o}\theta(\inv{P}\beta-\alpha)\theta(P\alpha-\beta)\\
+&(ic)^2 \delta_{P\alpha o}\delta_{\beta Pm}\theta(\beta-P\alpha)\theta(\alpha-\invbig{P}\beta)\\
-&(ic)^2\delta_{P\alpha\beta}\delta_\alpha m\theta(P\alpha-o)\theta(\invbig{P}o-\alpha)
\end{split}
\end{align}

The corresponding density function becomes
\begin{align}\label{at_1}
\begin{split}
&\rho_\downarrow(z)\\
=&\frac{\sigma}{2\sqrt{\pi}t}(e^{-\sigma^2(z-2tk_0)^2/4t^2}+2\sqrt{2} \pi t c^2(f(\frac{z}{2t})+g(\frac{z}{2t}))\\
&\times (h(N-\beta)+h(\beta-1))
\end{split}
\end{align}
with
 \begin{align}
 \begin{split}
&f(\xi)\\
=&\text{erfc}(-\frac{\sigma}{\sqrt{2}}(2\xi+\frac{2ct}{\sigma^2}-k_o\delta))e^{\frac{(k_0\sigma^2-2ct)^2}{2\sigma^2}+4ct\xi+c^2\sigma^2/2}
\end{split}
\end{align}
\begin{align}\begin{split}
&g(\xi)\\
=&\text{erfc}(\frac{\sigma}{\sqrt{2}}(2\xi-\frac{2ct}{\sigma^2}-k_o))e^{\frac{(k_0\sigma^2+2ct)^2}{2\sigma^2}-4ct\xi+c^2\sigma^2/2}
\end{split}
\end{align}
\begin{align}
h(m)=\frac{\text{exp}(-|c|a)-\text{exp}(-(m+1)|c|a)}{1-\text{exp}(-|c|a)}
\end{align}
Here we have dropped the terms when bra state and ket state have different pairs of bound state as such contribution is small by factor $e^{-a^2/8\sigma^2}$. Note, when t is large, both $f(\xi)$ and $g(\xi)$ diminishes as $\frac{\sigma}{ct}$, which cancels the $t$ in the prefactor of the second term in Eq. \eqref{at_1}. Moreover, as $h(m)\approx \text{exp}(-|c|a)$ for large $|c|a$, the bound state contribution is suppressed with strong interaction as $\sigma c e^{-|c|a}$.

Similarly, the density of the up spin can be obtained as
\begin{align}\label{at_1}
\begin{split}
&\rho_\uparrow(z)\\
=&\frac{\sigma}{2\sqrt{\pi}t}(e^{-\sigma^2(z-2tk_0)^2/4t^2}-2\sqrt{2} \pi t c^2(f(\frac{z}{2})+g(\frac{z}{2}))\\
&\times (h(N-\beta)+h(\beta-1))
\end{split}
\end{align}

The noise function becomes
\begin{align}
\begin{split}
&\tilde{C}(z,z',t)\\
=&C(z,z',t)+\frac{2\pi c^2 t}{N-1}
e^{\sigma^2(z-z'-k_0)^2/2-k_o^2\sigma^2/2-2 t|c(z-z')|}\\
&e^{c^2\sigma^2/2}(h(N-\beta)+h(\beta-1))+\frac{2\pi c^2 t}{N-1} e^{-\frac{\sigma^2 z'^2}{4t^2}}(f(\frac{z}{2t})\\
&+g(\frac{z}{2t})(h(N-\beta)+h(\beta-1))
\end{split}
\end{align}
with $C(z,z',t)$ being the noise function for the repulsive case, see Eq. \eqref{RNoise}. Note, unlike the density function, the long time behavior of the noise function close to the origin is dominated by bound states resulting from the second term in the above expression. The last term, however, is small in the large $t$ limit.

%%%%%%%%%%%%%%%%%%%%%%%%%%%%%%%%%%%%%%%%%%%%%%%%%%%%%%%%%%%%%%
\color{black}
\section{Bosonic Gaudin-Yang Model}
In this section, we will apply the Yudson approach to the quench problems in bosonic Gaudin-Yang system, and make a comparison of the results with those of the fermionic counterpart.

The bosonic Gaudin-Yang model is described by the same Hamiltonian as before, see Eq.~(\ref{HGY}), except that $\Psi_\sigma^\dagger(x)$ ($\Psi_\sigma(x)$) is bosonic creation (annihilation) operator satisfying commutation relation. The Bethe Ansatz eigenstate can be obtained as
\begin{align}\begin{split}
|\mu,k\rangle&=\sum_{\substack{P\in S_N\\R\in S_M}}\int_x \sum_\alpha e^{i\sum_i k_{Pi} y_i}\prod_{P_{ij}\in P}C^B(k_i-k_j)\prod_{m<n}\\
& S^B(\mu_m-\mu_n) \prod_m I^B(\mu_m, \alpha_m, Pk)\theta(x)\theta(\alpha)|x,\alpha\rangle
\end{split}\end{align}
with 
\begin{align}
C^B(k_i-k_j)=\frac{k_i-k_j-ic}{k_i-k_j+ic}
\end{align}
\begin{align}
S^B(\mu_m-\mu_n)=\frac{\mu_m-\mu_n-ic Sgn(\alpha_m-\alpha_n)}{\mu_m-\mu_n+ic}
\end{align}
\begin{align}
I^B(\mu,\alpha, k)=\frac{ic}{\mu-k_{\alpha}-ic/2}\prod_{n<\alpha}\frac{\mu-k_n+ic/2}{\mu-k_n-ic/2}
\end{align}
Here, the operator $P$, $R$, $\theta$-notation and state notation are defined in section~\ref{sec:GYM}. The Yudson state can be extracted  from it as
\begin{align}\begin{split}
|k,\mu)=\int_x \sum_\alpha e^{i\sum_i k_{Pi} y_i}\prod_m I^B(\mu_m, \alpha_m, k)\theta(x)\theta(\alpha)|x,\alpha\rangle
\end{split}\end{align}
Thus, the Yudson representation in real space becomes
\begin{align}
\begin{split}
&\int_C dk\int_{C'} d\mu \langle y,\alpha|k,\mu\rangle(k,\mu|x,\beta\rangle\theta(x)\theta(y)\theta(\alpha)\theta(\beta)
 \\=& \sum_{P,R} e^{i\sum_i k_i(y_{\inv{P}i}-x_i)}\prod_{P_{ij}\in P}C^B(k_i-k_j)\prod_{m<n}^M S^B(\mu_m\\
 &-\mu_n)\prod_{m=1}^M I^B(\mu_m,\alpha_m,Pk)I^{B*}(\mu_m,\beta_m,k)\theta(x)\theta(y)\\
 &\times\theta(\alpha)\theta(\beta)
\end{split}
\end{align}
The contour for $c>0$ is the same as that of the attractive case in the fermionic Gaudin-Yang model, i.e. the three-line contour, while the contour for $c<0$ duplicates the repulsive situation. To explain the choice of such contour, we now discuss how the central theorem can be proved by focusing on the aspects that are unique to the bosonic model.
\subsection{Central Theorem}

As we did for the fermionic case, we start with the single impurity situation. For both $c>0$ and $c<0$, the $\mu$ integral contour can be closed from above. This transforms the integration into pole contributions at $k_o-ic/2$ for any $o$ that satisfies the condition $o\leq\beta$ and $\invbig{P}o\geq\alpha$. The expressions of the residue here are the same as those in the fermionic case except for the sign in front of $c$. Thus, there are two type of poles. The one at $k_o=k_n$ is only apparent, due to the same reason as before. The other pole, which takes the form $k_i=k_j-ic$ with $i<\beta<j$ and $\invbig{P}i>\alpha>\invbig{P}j$ is a real one. At the same time, it must be true that $P_{ij}\in P$, therefore, the denominator is canceled by the numerator in $C^B(k_i-k_j)$, leaving only Lieb-Liniger type of poles as defined on page~\pageref{lltype}. Following the argument in~\cite{deepak1}, one can see that these poles do not contribute to the wavefunction integration. 

In the presence of multiple $\mu$'s, one should carry our the integration over $\mu_1,\ldots,\mu_M$ repeated by closing each contour in the upper half plane. Using the same argument as we made in section~\ref{sec:mltyCT}, one can show that the collection of poles lying above the integral contour of $\mu_m$ is unaffected  by the product of scattering matrices among $\mu$'s, as long as we do the integration in the aforementioned order. Then each $\mu_m$ integration results in a collection of poles of the form $k_i=k_j-ic$ for $i<\beta_m<j$ and $\invbig{P}i>\alpha_m>\invbig{P}j$. At the same time, $\mu_m$ is related to $k_i$ as $\mu_m=k_i-ic/2$. As no two $\mu$'s can be identical in the nested Bethe Ansatz, the $\mu$ integration cannot take the residue at the same point. This guarantees that each pole appears at most once in the denominator. This pole will then be cancelled by the scattering matrices among $k$'s. Thus, we still have only Lieb-Liniger type of poles. Further proof of the central theorem in section~\ref{sec:ct} can be applied directly to here, and one also gets the same normalization constant.
\subsection{Bound states}\label{sec:BGY_boundstates}
In the previous part, we have shown that all poles comes from $C^B(k_i-k_j)$. What this indicates is twofold. First, the $k-\mu$ strings (type 2) disappear. Secondly, $k$ strings (type 5) emerge. That means the formation of bound states no longer depends on the existence of a spinor. This makes sense as Lieb-linger gases bind together with attractive interaction too. Now, there are three types of strings, $k-\mu$ pairs (type 1), $k$ strings (type 5) and composites of these two.
A composite may be formed if a $k$ in the $k-\mu$ pairs coexists in the $k$ strings which snaps the two string together, see Figure~\ref{fig:stringtype}.(6). Note, although the $k-\mu$ string is not a basic type, it may emerge has a composite. For a complete set of basis, one do need to include all composite configuration as well as the $k-\mu$ pair and $k$ strings. This makes the enumeration more complicated. Physically, this is due to the fact that bosonic wavefunction are symmetric, thus particles have more overlap in the highly polarized limit. Therefore, there are more interaction among bosons than fermions. 

\subsection{Time Evolution}
When the system has only one spin-up boson and one spin-down boson, the wave function for time evolved state is identical to that of the fermionic counterpart. This is because the quantum nature of particles only affects the symmetry property among identical particles. When the number of the majority bosons get greater, we do expect to see different behavior between bosonic and fermionic systems. However, our calculations of the density and noise correlation for the repulsive case in leading order of $\sigma$ yield the same results in both scenarios. The sign difference in front of $c$ is canceled by symmetry difference under exchanging two particles. Like in the previous model, $c(0,0,t)\to -1$ for large time, indicating that the particles develop a trend to avoid overlap with each other for our chosen initial state. We relate this phenomenon to energy conservation which plays the same role in both systems.

Note, dropping higher order terms in $\sigma$ as we did in Eq. \ref{drop} decouples the two measured degrees of freedom from the rest of the system. Making the multi-particle problem($N>2$) equivalent to one with two distinguishable particles. This is not the case for a system with bound states. In these states, one of the measured particle binds together with a third particle. Thus, two of the particles become indistinguishable and statistics plays a big role. We believe that the quench dynamics will be greatly different from the fermionic counterpart, if the bound states contribute significantly, i.e. in an attractive system with a lot of overlap in the initial state. We shall study these issues  in later publications

\section{Conclusions}
In this work  the Yudson approach has been introduced as an eigenstate expansion of a general state that is well separated in the coordinate space. We have applied this method to the study of two-component Fermi (Bose) gases. We have specified the integral contour for any spin imbalance with either attractive or repulsive interaction. We have proved the central theorem which shows why such contour is chosen. We have shown that the contour integral implicitly includes all free and bound states. By separating these states apart, we have enumerated all bound state solutions. Some of them are predicted by the String hypothesis, while others are not. This provides us with a complete basis in the Hilbert space that is greater than long believed. We have also discussed the formation mechanism for these bound states, some of which are of FFLO type. The time evolution has also been addressed for one impurity problem. Exact wavefunction has been obtained for two distinct fermions. We have seen similar density profile for either type of interaction and particle. This is due to the fact that the fermions (bosons) have negligible overlaps in the initial state and are refrained from contact to avoid change of energy. This picture is confirmed by the calculation of noise function. Asymptotic behavior of more particles has been studied.  We noticed that bound state contribution to density is suppressed by factor $c\sigma e^{-|c|a}$, while for the noise function, the bound state contribution comes with the fact $c^2 t e^{-|c|a}$. That is to say that although the bound state can be barely seen in the density profile with lattice initial state, its effect dominates in the noise function in the vicinity of the origin asymptotically.

We have not studied the quench dynamics of $M$ ($M>1$) impurities, which is an interesting problem which we will leave for future study. The quench dynamics from an initial state with prominent overlaps has not be considered and is not easily solved with the Yudson approach. Integration with all the Heaviside step functions is difficult, if possible, to perform. The combination of the form factor approach and the Yudson approach is a promising direction. A key ingredient is an explicit form factor that works for both diagonal and off-diagonal elements. They should also work well for complex parameters. This is what is missing so far. Such form factors free one from evaluating high dimensional integrals of spatial coordinates, which we have to make approximation about. Last but not least, it is desirable if one could work out the $k$ and $\mu$ integrations without exploiting the saddle point approximation. If one changes the variable to include the imaginary part of the contour, the Yusdon representation is nothing but a real integration. And one no longer needs to include different bound state contribution separately. 
\section{Acknowledgements}
H. Guan and N. Andrei were supported by NSF Grant DMR 1410583.

\normalem
\bibliography{thesis}

\end{document}